\documentclass[tikz,11pt,english,a4paper]{article}
\usepackage{float}
\usepackage{jcappub}
\usepackage[utf8]{inputenc}

\usepackage{fancyvrb}
\usepackage{keyval}

\usepackage{bm}
\usepackage{algorithm}
\usepackage{nicefrac}
\usepackage{bbold}
\usepackage{dsfont}
\usepackage{setspace}
\usepackage[linewidth=1pt]{mdframed}
\usepackage{bold-extra}
\usepackage{tabto}
\usepackage{amsmath}
\usepackage{cleveref}
\crefname{equation}{eq.}{eqs.}
\Crefname{equation}{Eq.}{Eqs.}

\crefname{figure}{figure}{figures}
\Crefname{figure}{Figure}{Figures}

\crefname{table}{table}{tables}
\Crefname{table}{Table}{Tables}

\crefname{section}{section}{sections}
\Crefname{section}{Section}{Sections}

\crefname{appendix}{appendix}{appendices}
\Crefname{appendix}{Appendix}{Appendices}

\usepackage{mathrsfs}

\usetikzlibrary{angles, quotes}
\usetikzlibrary{decorations.markings, arrows.meta}
\tikzset{
  fermion/.style={
    decoration={markings, mark=at position 0.5 with {\arrow{Stealth}}},
    postaction={decorate}
  }
}

\usepackage{algorithm}
\usepackage[noend]{algpseudocode}
\makeatletter
\def\BState{\State\hskip-\ALG@thistlm}
\makeatother

\usepackage{fontawesome5}
\usepackage{hyperref}
\makeatletter
\newcommand{\github}[1]{%
   \href{#1}{\faGithub}%
}
\makeatother
\newcommand{\integralnotebook}{\href{https://gist.github.com/086ba145090d94d70b18ca4ef3617a36}{\texttt{IntegralComputation.ipynb}}}

\newcommand{\class}{\textsc{class}}
\newcommand{\camb}{\textsc{camb}}

\usepackage{siunitx}
\DeclareSIUnit \parsec {pc}

\usepackage[T1]{fontenc}

\usepackage{lmodern} 
\rmfamily 
\DeclareFontShape{T1}{lmr}{b}{sc}{<->ssub*cmr/bx/sc}{}
\DeclareFontShape{T1}{lmr}{bx}{sc}{<->ssub*cmr/bx/sc}{}

\usepackage{soul}

\usepackage[edges]{forest}
\definecolor{folderbg}{RGB}{124,166,198}
\definecolor{folderborder}{RGB}{110,144,169}
\newlength\Size
\tikzset{%
  folder/.pic={%
    \filldraw [draw=folderborder, top color=folderbg!50, bottom color=folderbg] (-1.05*\Size,0.2\Size+5pt) rectangle ++(.75*\Size,-0.2\Size-5pt);
    \filldraw [draw=folderborder, top color=folderbg!50, bottom color=folderbg] (-1.15*\Size,-\Size) rectangle (1.15*\Size,\Size);
  },
  file/.pic={%
    \filldraw [draw=folderborder, top color=folderbg!5, bottom color=folderbg!10] (-\Size,.4*\Size+5pt) coordinate (a) |- (\Size,-1.2*\Size) coordinate (b) -- ++(0,1.6*\Size) coordinate (c) -- ++(-5pt,5pt) coordinate (d) -- cycle (d) |- (c) ;
  },
}
\forestset{%
  declare autowrapped toks={pic me}{},
  declare boolean register={pic root},
  pic root=0,
  pic dir tree/.style={%
    for tree={%
      folder,
      font=\ttfamily,
      grow'=0,
    },
    before typesetting nodes={%
      for tree={%
        edge label+/.option={pic me},
      },
      if pic root={
        tikz+={
          \pic at ([xshift=\Size].west) {folder};
        },
        align={l}
      }{},
    },
  },
  pic me set/.code n args=2{%
    \forestset{%
      #1/.style={%
        inner xsep=2\Size,
        pic me={pic {#2}},
      }
    }
  },
  pic me set={directory}{folder},
  pic me set={file}{file},
}

\usepackage{newfloat}
\usepackage{caption}
\DeclareFloatingEnvironment[fileext=frm,placement={!t},name=Algorithm]{pseudoenv}
\DeclareCaptionLabelSeparator{normal-colon}{\normalfont : } 
\makeatletter
\g@addto@macro\bfseries{\boldmath}
\makeatother

\makeatletter
\def\old@comma{,}
\catcode`\,=13
\def,{%
  \ifmmode%
    \old@comma\discretionary{}{}{}%
  \else%
    \old@comma%
  \fi%
}
\makeatother

\newcounter{Meq}

\newcounter{Leq}

\begin{document}


\title{Self-interacting neutrinos in cosmological perturbation theory - integrating the collision kernel}

\author[a]{Jakob Krukow Mogensen,}
\author[a]{Steen Hannestad, }
\author[a]{Thomas Tram,}
\affiliation[a]{Department of Physics and Astronomy, Aarhus University,
 DK-8000 Aarhus C, Denmark}

\emailAdd{jakob@mogensen.me}
\emailAdd{steen@phys.au.dk}
\emailAdd{thomas.tram@phys.au.dk}

\abstract{
Cosmological constraints on self-interacting neutrinos require a Boltzmann hierarchy in which the collision term is projected onto momentum-averaged multipoles.  We revisit the collision kernel for neutrino-neutrino scattering mediated by a light scalar and derive an exact analytic expression for the multipole integral that determines the coefficients $\alpha_\ell$.  The key idea is to express the integration kernel as angular derivatives of the Yukawa-potential $\frac{\mathrm{e}^{-P/2}}{P}$, move the derivatives onto Legendre polynomials, and reduce the remaining momentum integrals to a single base family obeying a first-order recurrence.  This gives an exact rational-plus-$\pi^2$ representation for every multipole, together with a compact implementation based on exact rational arithmetic.  We provide the recurrence relations, a closed form for the base integral, and an asymptotically constrained approximation suitable for Boltzmann codes such as \class{}. Our numerical implementation is publicly available in the Jupyter notebook \integralnotebook.
}

\maketitle

\section{Introduction}\label{sec:introduction}

Within the standard model of particle physics neutrinos are strictly massless particles. However, it has long been known from both astrophysical and terrestrial neutrino oscillation observations that neutrinos do indeed possess non-zero masses (see e.g. \cite{Fukuda_1998,Ahmad_2002}  as well as \cite{Esteban:2024eli} for a recent review).
The fact that neutrino masses are tiny relative to the masses of other standard model fermions is very likely related to the fact that neutrinos are chargeless, allowing for a Majorana mass term and the possibility of tiny neutrino masses via the See-Saw mechanism (see e.g.\ \cite{Mohapatra:2006gs}).

In any case, non-zero neutrino masses clearly point to neutrinos possessing interactions with new particle degrees of freedom beyond the standard model, usually referred to as non-standard or secret neutrino interactions (see \cite{Berryman:2022hds} for a recent, comprehensive review). Given that neutrinos are exceedingly hard to probe in laboratory experiments, it is no surprise that many of the tightest constraints on these new interactions come from astrophysical or cosmological considerations.

In cosmology, it has long been known that non-standard neutrino interactions can dramatically alter observables such as the cosmic microwave background anisotropy and properties of the cosmological large scale structure. This is possible because  such interactions can prevent free-streaming of neutrinos provided that they are sufficiently strong 
(see \cite{Hannestad_2005,Bell:2005dr,Friedland:2007vv} for early works on this). 
The cosmological impact of neutrino non-standard interactions in general depends strongly on specifics of the interaction, and many possible scenarios have been investigated \cite{Berryman:2022hds}. 
However, the simplest scenarios typically involve neutrinos coupling to a new scalar or pseudo-scalar degree of freedom with a Lagrangian of the form 
\begin{gather}
  \mathcal{L}_\mathrm{int} = \mathfrak{g}_{ij} \bar{\nu}_i \nu_j \phi + \mathfrak{h}_{ij} \bar{\nu}_i \gamma_5 \nu_j \phi, \label{eq:majoron_lagrangian},
\end{gather}
as would be the case in e.g.\ majoron models of neutrino mass generation ~\cite{Chikashige:1980ui,Gelmini:1980re}.

The cosmological phenomenology of such new interactions depends strongly on whether $\phi$ is massless or massive. 
In the case of a massless $\phi$ it is possible to produce physical $\phi$ particles even at low temperature, either by decay $\nu_i \to \nu_j \phi$, or by pair production $\nu \bar\nu \to \phi \phi$. This means that the new pseudoscalar\footnote{In the case of interactions with a new massless particle, the interaction would most likely be pseudo-scalar in nature to avoid possible effects of a new long range 5\textsuperscript{th} force in the neutrino sector.} must be kept track of separately when computing for example cosmological perturbations. Phenomenologically this model can also lead to what have been called the neutrinoless universe \cite{Beacom:2004yd} in which neutrinos pair annihilate away as soon as the temperature drops below their rest mass, leaving only massless pseudoscalars behind.

Here, we will focus on the somewhat simpler case where $\phi$ is assumed to be very massive such that any primordial population of these particles can be safely assumed to have been removed by Boltzmann suppression. This model is particularly simple in the sense that the only cosmological effect is to introduce a new coupling term between neutrinos, i.e.\ a non-zero right-hand side of the Boltzmann equation. This model has been studied numerous times in cosmology and found to have interesting properties, most intriguingly it has been found that a fairly strong coupling is not necessarily excluded by current data (for a non-exhaustive list see 
\cite{Archidiacono:2013dua,Cyr_Racine_2014,Kreisch:2019yzn,RoyChoudhury:2020dmd,Pal:2026cgj,Perez-Castro:2026muj,Montefalcone:2025ibh,Poudou:2025qcx,Camarena:2024daj}).
\\~\\
Almost all treatments have used the Relaxation Time Approximation (RTA) first introduced in \cite{Hannestad:2000gt} where the very complex collision term is replaced by an effective momentum-averaged quantity encoding the interaction rate of the system.
The validity of this approximation was carefully tested in \cite{Oldengott_2015,Oldengott_2017} and found to be valid to excellent precision provided that a set of precomputed numerical front factors are used. The computation of these numerical factors has previously involved solving a highly unstable triple integral. However, in the present work we demonstrate that it is possible to carry out the entire integration procedure analytically.
\\~\\
The paper is structured as follows: \Cref{sec:theory} contains a brief outline of cosmological perturbation theory with self-interacting neutrinos, \cref{sec:boltzmantomaster} describes how to arrive at the integral expression for the numerical front factors to be used in the relaxation time approximation, and \cref{sec:recursion} outlines the procedure for carrying out the analytic integration. \Cref{sec:observables} contains a brief summary of numerical results, and finally, \cref{sec:discussion} contains our conclusions. Details pertaining to the analytic integration procedure can be found in appendices~\ref{app:base_integral}-\ref{app:implementation_details}.

\section{Theory}\label{sec:theory}
In this section, we will briefly review the derivation of the scattering term and the Boltzmann hierarchy. We refer the reader to references~\cite{Ma_1995,Hannestad_2005,Oldengott_2015,Oldengott_2017,Dodelson_2020} for the details.

\subsection{The Boltzmann Equation}
Our starting point is the relativistic Boltzmann equation for the phase-space distribution $f(\mathbf{x}, \mathbf{P}, \tau)$ at position $\mathbf{x}$, three-momentum $\mathbf{P}$, and conformal time $\tau$,
\begin{gather}
  \left(\frac{\partial f}{\partial \tau}\right)_\mathrm{coll} = P^\alpha \frac{\partial f}{\partial x^\alpha} - \Gamma^\gamma_{\alpha\beta} P^\alpha P^\beta \frac{\partial f}{\partial P^\gamma}, \label{eq:boltzmann_relativistic}
\end{gather}
where Einstein notation is adopted, $\Gamma^\gamma_{\alpha\beta}$ denotes the Christoffel symbols, and $P^\alpha$ is the canonical four-momentum conjugate to the comoving four-coordinate $x^\alpha$.
The collision term encapsulates all the relevant scattering processes for the particle species in focus.
As such, utilising the notation of ref.~\cite{Ma_1995}, we choose to work in the synchronous gauge, where the infinitesimal line element $\mathrm{d}s$ is perturbed via
\begin{gather}
  \mathrm{d}s^2 = a^2(\tau) \left(-\mathrm{d}\tau^2 + (\delta_{ij} + h_{ij}) \,\mathrm{d}x^i \,\mathrm{d}x^j\right). \label{eq:synchronous_metric}
\end{gather}
Here, $a$ denotes the scale factor, $\delta_{ij}$ Kronecker's delta, and $h_{ij}$ the metric perturbation.
We will express any momentum in terms of the comoving momentum $\mathbf{q} \equiv a\mathbf{p}$, and any energy in terms of the comoving energy $\epsilon \equiv aE$.
Likewise, we split the phase-space distribution function of a species $i$ into an unperturbed background $\bar{f}$ and a perturbation contribution $\Psi$,
\begin{gather}
  f_i(\mathbf{x}, \mathbf{q}, \tau) = \bar{f}_i(q, \tau) \left[ 1 + \Psi_i(\mathbf{x}, \mathbf{q}, \tau) \right]. \label{eq:distribution_splitting}
\end{gather}
$q \equiv |\mathbf{q}|$ is used for the background, as it relies on the assumption of it being homogeneous and isotropic, hence the lack of position- and direction dependence.
Since the neutrinos in question interact, $\bar{f}$ must be calculated using the homogeneous limit of the Boltzmann equation~\eqref{eq:boltzmann_relativistic}, i.e.
\begin{gather}
  \dot{\bar{f}}_i(q, \tau) = \left( \frac{\partial f_i}{\partial \tau} \right)^{(0)}_\mathrm {coll}, \label{eq:background_collision}
\end{gather}
where the overdot denotes a derivative with respect to conformal time. Now, a linear expansion of \cref{eq:boltzmann_relativistic} in the perturbed quantities yields an equation of motion for the perturbation contribution of \cref{eq:distribution_splitting}. In Fourier space, it takes the form
\begin{gather}
  \frac{1}{\bar{f}_i} \left(\frac{\partial f_i}{\partial \tau}\right)^{(1)}_\mathrm{coll}
  = \dot{\Psi}_i(\mathbf{k}, \mathbf{q}, \tau)
  + \mathrm{i}\frac{qk}{\epsilon}(\hat{k}\cdot\hat{q})\Psi_i(\mathbf{k}, \mathbf{q}, \tau)
  + \frac{\partial \ln \bar{f}_i(q, \tau)}{\partial \ln q} \left(\dot{\eta}
  - (\hat{k}\cdot\hat{q})^2\frac{\dot{h}+6\dot{\eta}}{2}\right), \label{eq:perturbation_e.o.m.}
\end{gather}
where $h \equiv h^i_{\;i}(\mathbf{k}, \tau)$, and
\begin{gather}
  6\eta \equiv 6\eta(\mathbf{k}, \tau) = -\frac{3}{2k^4} k^i k^j h_{ij} + \frac{1}{2k^2}h
\end{gather}
are the Fourier transforms of, respectively, the trace and traceless space-space longitudinal perturbations of the synchronous metric~\eqref{eq:synchronous_metric}.
The two collision terms, \cref{eq:background_collision,eq:perturbation_e.o.m.}, make up the baseline of what becomes the Boltzmann hierarchy.
We will adopt the tetrad basis: an orthonormal basis in which an observer is comoving with the synchronous line element in \cref{eq:synchronous_metric}.
Here, considering a $CP$ invariant two-body scattering process $i(p^\alpha) + j(n^\alpha) \rightarrow k(p'^\alpha) + l(n'^\alpha)$, the collision term is given by
\begin{align}
  \left(\frac{\partial f_i}{\partial t}\right)_\mathrm{coll}(\mathbf{k}, \mathbf{p}, \tau)
  &= \frac{a}{2E(\mathbf{p})} \int (2\pi)^4 \delta^{(4)}_\mathrm{D} (p^\alpha+n^\alpha-p'^\alpha-n'^\alpha) |\mathcal{M}_{ij \rightarrow kl}|^2 \notag\\
  &\qquad \times \Big(f_k(\mathbf{k}, p', \tau)f_l(\mathbf{k}, n', \tau) \big[1 \pm f_i(\mathbf{k}, p, \tau)\big] \big[1 \pm f_j(\mathbf{k}, n, \tau)\big] \notag\\
  &\qquad\qquad -f_i(\mathbf{k}, p, \tau)f_j(\mathbf{k}, n, \tau) \big[1 \pm f_k(\mathbf{k}, p', \tau)\big] \big[1 \pm f_l(\mathbf{k}, n', \tau)\big] \Big) \notag\\
  &\qquad\qquad\qquad \mathrm{d}^3\mathbf{\Pi}_j(\mathbf{n}) \,\mathrm{d}^3\mathbf{\Pi}_k(\mathbf{p}') \,\mathrm{d}^3\mathbf{\Pi}_l(\mathbf{n}'), \label{eq:collision_term}
\end{align}
where
\begin{gather}
  \mathrm{d}^3\mathbf{\Pi}_a(\mathbf{n}) \equiv g_a \frac{\mathrm{d}^3\mathbf{n}}{(2\pi)^3 2E(\mathbf{n})} \label{eq:differential_momentum_per_energy}
\end{gather}
with $g_a$ being the number of internal degrees of freedom of particle species $a\in\{j, k, l\}$. $p^\alpha$, $n^\alpha$, etc. denote four-momenta, $\delta^{(4)}_\mathrm{D}$ is the four-dimensional Dirac delta function, and $|\mathcal{M}_{ij \rightarrow kl}|^2$ is the Lorentz-invariant norm-squared matrix element that is to be determined by quantum field theory.
This is averaged over the final particle spins and includes a symmetrisation factor $1/(N_i!N_f!)$ that depends on the number of particles in the initial and final states.

To reduce the complexity, we will neglect the effects of Bose enhancement and Pauli blocking and instead enforce Maxwell-Boltzmann statistics.
This means that we set $1-f_\nu\sim1$ and $1+f_\phi\sim1$. When imposing the distribution function splitting, \cref{eq:distribution_splitting}, the collision integral for the zeroth-order phase-space distribution function takes the form
\begin{align}
  \left( \frac{\partial f_i}{\partial \tau} \right)^{(0)}_{ij \leftrightarrow kl}(q, \tau)
  &= \frac{g_j g_k g_l}{2q(2\pi)^5} \int \delta_\mathrm{D}^{(4)}(q^\alpha+l^\alpha-q'^\alpha-l'^\alpha) |\mathcal{M}_{ij \leftrightarrow kl}|^2 \notag\\
  &\qquad\times \left[ \bar{f}_k(q', \tau) \bar{f}_l(l', \tau) - \bar{f}_i(q, \tau) \bar{f}_j(l, \tau) \right] \frac{\mathrm{d}^3\mathbf{q'}}{2q'} \frac{\mathrm{d}^3\mathbf{l}}{2l} \frac{\mathrm{d}^3\mathbf{l'}}{2l'}.
\end{align}
Similarly, the time evolution of the first-order phase-space distribution reads
\begin{align}
  &\left( \frac{\partial f_i}{\partial \tau} \right)_{ij \leftrightarrow kl}^{(1)}(\mathbf{k}, \mathbf{q}, \tau)
  = \frac{g_j g_k g_l}{2q(2\pi)^5} \int \delta_\mathrm{D}^{(4)}(q^\alpha+l^\alpha-q'^\alpha-l'^\alpha) |\mathcal{M}_{ij \leftrightarrow kl}|^2 \notag\\
  &\qquad\times \Big( \bar{f}_k(q', \tau) \bar{f}_l(l', \tau) \Psi_l(\mathbf{k}, \mathbf{l}', \tau)
  + \bar{f}_l(l', \tau) \bar{f}_k(q', \tau) \Psi_k(\mathbf{k}, \mathbf{q}', \tau) \notag\\
  &\qquad\qquad - \bar{f}_i(q, \tau) \bar{f}_j(l, \tau) \Psi_j(\mathbf{k}, \mathbf{l}, \tau)
  + \bar{f}_j(l, \tau) \bar{f}_i(q, \tau) \Psi_i(\mathbf{k}, \mathbf{q}, \tau) \Big) \frac{\mathrm{d}^3\mathbf{q}'}{2q'} \frac{\mathrm{d}^3\mathbf{l}}{2l} \frac{\mathrm{d}^3\mathbf{l}'}{2l'}.
\end{align}
The solutions of these two equations are outlined below in the limit where the mass of the scalar mediator far exceeds the typical energies of the neutrinos.

\subsection{The massive scalar limit}
\begin{figure}[tb]
  \centering
  \begin{tikzpicture}

    \begin{scope}[xshift=0cm]
      \coordinate (a) at (0, 0);
      \coordinate (b) at (2, 0);
      \draw[fermion] (-0.8, 1.8) node[left] {$\nu$} -- (a);
      \draw[fermion] (-0.8,-1.8) node[left] {$\nu$} -- (a);
      \draw[dashed] (a) -- node[above] {$\phi$} (b);
      \draw[fermion] (b) -- (2.8, 1.8) node[right] {$\nu$};
      \draw[fermion] (b) -- (2.8,-1.8) node[right] {$\nu$};
    \end{scope}

    \begin{scope}[xshift=6cm]
      \coordinate (a) at (0, 0.8);
      \coordinate (b) at (0,-0.8);
      \draw[fermion] (-1.4, 1.8) node[left] {$\nu$} -- (a);
      \draw[fermion] (-1.4,-1.8) node[left] {$\nu$} -- (b);
      \draw[dashed] (a) -- node[right] {$\phi$} (b);
      \draw[fermion] (a) -- ( 1.4, 1.8) node[right] {$\nu$};
      \draw[fermion] (b) -- ( 1.4,-1.8) node[right] {$\nu$};
    \end{scope}

    \begin{scope}[xshift=10.6cm]
      \coordinate (a) at (0, 0.8);
      \coordinate (b) at (0,-0.8);
      \draw[fermion] (-1.4, 1.8) node[left] {$\nu$} -- (a);
      \draw[fermion] (-1.4,-1.8) node[left] {$\nu$} -- (b);
      \draw[dashed] (a) -- node[left] {$\phi$} (b);
      \draw[fermion] (b) -- ( 1.4, 1.8) node[right] {$\nu$};
      \draw[fermion] (a) -- ( 1.4,-1.8) node[right] {$\nu$};
    \end{scope}

  \end{tikzpicture}
  \caption{Tree-level diagrams of the $\nu\nu \rightarrow \nu\nu$ neutrino--neutrino scattering via a massive scalar mediator $\phi$ in the $s$, $t$, and $u$ channels. Described by the Lagrangian~\eqref{eq:majoron_lagrangian}.}
  \label{fig:neutrino_stu_diagrams}
\end{figure}
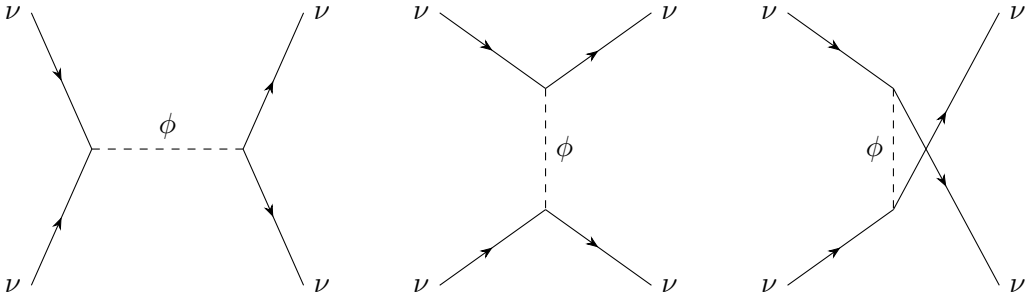

\begin{figure}[tb]
    \centering
    \includegraphics[width=\columnwidth]{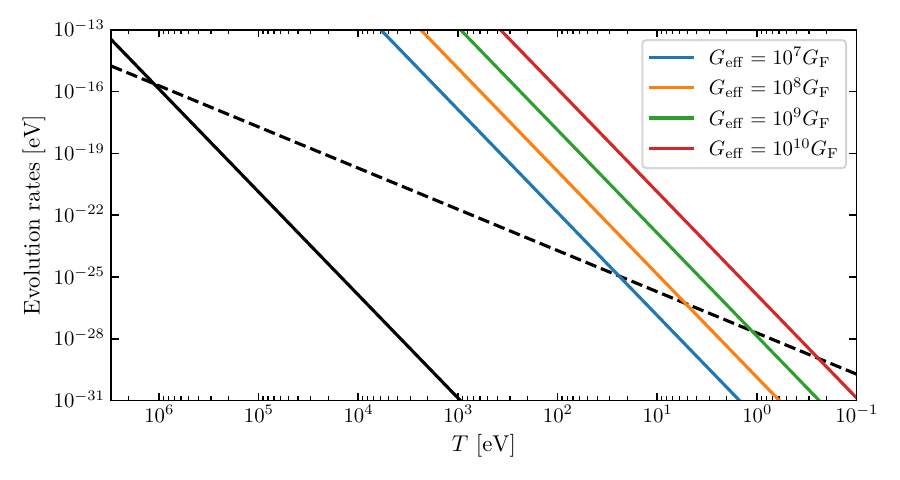}
    \caption{
      Comparison of different evolution rates.
      The colored lines show the self-interacting Majorana neutrinos.
      As for the black lines, the solid shows the standard weak interaction rate, while the dashed shows the Hubble rate in the radiation-dominated epoch.
      The decoupling from the cosmic plasma happens when the interaction rate drops below the Hubble rate.
    }
    \label{fig:neutrino_decouplings}
\end{figure}

Equipped with \crefrange{eq:background_collision}{eq:differential_momentum_per_energy}, one is indeed in a position to start tackling the collision term~\eqref{eq:collision_term}.
However, a number of complications quickly arise, as the matrix element, being $m_\phi$-dependent, comes with a complex angular dependence.
Therefore, we focus on the massive mediator case.
Indeed, having such a mediator implies, since lifetime roughly is determined by the inverse mass, that any initial population of the mediator will annihilate or decay into neutrinos way before any times of interest, leaving the interactions $\nu\phi\rightarrow\nu\phi$ and $\phi\phi\rightarrow\nu\nu$ irrelevant.
It also implies that the interaction $\nu\nu\rightarrow\phi\phi$ is kinematically suppressed and therefore not to be considered as well.
\\~\\
Motivated by the fact that scalar and pseudoscalar interactions are phenomenologically identical in cosmological settings, we set $\mathfrak{h}_{ij}=0$ in \cref{eq:majoron_lagrangian}.
Finally, we impose the ansatz that the couplings are flavour-independent and diagonal, i.e. $\mathfrak{g}_{ij}=\mathfrak{g}\delta_{ij}$.
Thus, we turn our attention to the scattering process $\nu\nu\rightarrow\nu\nu$ only.
As we are dealing with identical fermions, the tree-level diagrams described by the Lagrangian~\eqref{eq:majoron_lagrangian} include all three Mandelstam channels, as illustrated in \cref{fig:neutrino_stu_diagrams}.
In this framework, neutrinos still decouple from the rest of the cosmic plasma at the weak decoupling temperature, but remain scattering at a rate per particle
\begin{gather}
  \Gamma \sim \mathfrak{g}^4T_\nu^5/m_\phi^4 \equiv G_\mathrm{eff}^2T_\nu^5. \label{eq:scattering_rate}
\end{gather}
Note that the neutrino temperature dependence is the same as with weak interactions, such that the self-interactions only serve to \textit{delay} the epoch of neutrino decoupling\footnote{This relies on the assumption that the background phase space equilibrium distribution is unaffected by the neutrino self-interaction, even after it drops below the Hubble rate. See discussion in section 4.1 of ref.~\cite{Oldengott_2015}.}.
This is illustrated in \cref{fig:neutrino_decouplings}.
\noindent The condition $G_\mathrm{eff}^2 \gg G_F^2$, with $G_F$ being the Fermi constant, is a crucial assumption for this work, as it allows the definition of the rate of change of the neutrino opacity \cite{Cyr_Racine_2014}
\begin{gather}
  \dot{\kappa}_\nu \equiv -a G_\mathrm{eff}^2 T_\nu^5 \label{eq:neutrino_opacity}.
\end{gather}
This parameter will later feature in the Boltzmann hierarchy and come in handy when deriving an explicit expression for the collision kernel.

\subsection{Matrix element and collision integral reduction}
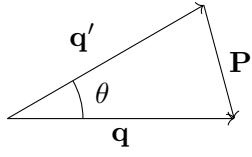
\begin{figure}[tb]
  \centering
  \begingroup
  \catcode`\,=12
  \begin{tikzpicture}

    \coordinate (O)  at (0,0);
    \coordinate (Q)  at (3,0);
    \coordinate (Qp) at ({3*0.866},{3*0.5});

    \draw[->] (O)  -- (Q)  node[midway, below] {$\mathbf{q}$};
    \draw[->] (O)  -- (Qp) node[midway, above left] {$\mathbf{q}'$};
    \draw[->] (Qp) -- (Q)  node[midway, right] {$\mathbf{P}$};

    \draw (1,0) arc[start angle=0, end angle=30, radius=1cm];
    \node at ({1.3*cos(15)},{1.3*sin(15)}) {$\theta$};

  \end{tikzpicture}
  \endgroup
  \caption{Vectors $\mathbf{q}$, $\mathbf{q}'$, and $\mathbf{P} = \mathbf{q} - \mathbf{q}'$. The angle between $\mathbf{q}$ and $\mathbf{q}'$ is denoted $\theta$.}
  \label{fig:momentum_vectors}
\end{figure}

Since the neutrinos scatter by means of a heavy scalar mediator, any preexisting helicity state asymmetries will have evolved to zero long before the relevant neutrino perturbations enter the horizon. This means that we may assume identical phase space distributions for the two initial neutrino helicity states, simplifying the averaging of the norm-squared matrix element $|\mathcal{M}_{ij \rightarrow kl}|^2$ in \cref{eq:collision_term}.
The result is\footnote{We note that in the limit of interest here, i.e.\ massless Majorana neutrinos interacting via exchange of a very massive mediator, leads to the same matrix element structure (up to constants front factors) for both scalar, pseudo-scalar, and axial vector coupling. Vector and derivative couplings yield a vanishing matrix element in the exactly massless limit. Our calculation here is therefore valid for any of the mentioned types of mediators.}
\begin{gather}
  |\mathcal{M}_{ij \rightarrow kl}|^2 = \frac{\mathfrak{g}^4}{2m_\phi^4} (s^2 + t^2 + u^2),
\end{gather}
where the Mandelstam variables are defined as
\begin{align}
  s &\equiv (p^\alpha + n^\alpha)^2 = (p'^\alpha + n'^\alpha)^2, \\
  t &\equiv (p^\alpha - p'^\alpha)^2 = (n^\alpha - n'^\alpha)^2, \\
  u &\equiv (p^\alpha - n'^\alpha)^2 = (n^\alpha - p'^\alpha)^2.
\end{align}
What is left to evaluate of \cref{eq:collision_term} are the collision integrals.
Under the assumption of Maxwell-Boltzmann statistics, the neutrino background distribution reads
\begin{gather}
  \bar{f}_\nu(q) = \mathrm{N} \,\mathrm{exp} \left( -\frac{q}{aT_\nu} \right) = \mathrm{N} \,\mathrm{exp} \left( -\frac{q}{T_{\nu,0}} \right), \label{eq:neutrino_background}
\end{gather}
where $T_{\nu,0} = 1.95 \,\mathrm{K}$ is the present-day neutrino temperature.
We may use this temperature because the interaction $\nu\nu\leftrightarrow\nu\nu$ is strictly confined within its own sector, meaning that $e^-e^+$-entropy transfer has the same footprint as the standard case on the post-annihilation neutrino-photon temperature relation.
$\mathrm{N} = 7\pi^4/720$ is a normalisation factor introduced to make $a^{-3} \int \bar{f}_\nu(q) \,\mathrm{d}^3\mathbf{q}$ match the relativistic Fermi-Dirac energy density.
This means that, when inserting the distribution function splitting, \cref{eq:distribution_splitting}, into the right-hand side of the collision term, \cref{eq:collision_term}, the time evolution of the zeroth-order phase-space distribution term vanishes,
\begin{gather}
  \left( \frac{\partial f_\nu}{\partial \tau} \right)_{\nu\nu\leftrightarrow\nu\nu}^{(0)}(q, \tau) = 0.
\end{gather}
In the massive scalar limit, where we only need to consider the process $\nu\nu\leftrightarrow\nu\nu$, the first-order distribution evaluates to
\begin{align}
  &\left( \frac{\partial f_\nu}{\partial \tau} \right)_{\nu\nu\leftrightarrow\nu\nu}^{(1)}(\mathbf{k}, \mathbf{q}, \tau)
  = -\frac{2\mathrm{N}\mathfrak{g}^4}{a^4m_\phi^4(2\pi)^3} \bigg( \frac{80}{3}qT_{\nu,0}^4 \bar{f}_\nu(q, \tau) \Psi_\nu(\mathbf{k}, \mathbf{q}, \tau) \notag\\
  &\qquad + \int \frac{q'}{q} \left[ \frac{5q^2q'^2}{6} \mathrm{e}^{-q/T_{\nu,0}} (1-\mu)^2 - K(q, q', \mu) \right] \bar{f}(q', \tau) \Psi_\nu(\mathbf{k}, \mathbf{q}', \tau) \,\mathrm{d}\mu \,\mathrm{d}q' \bigg), \label{eq:firstorder_massive_collision}
\end{align}
where the integration kernel\footnote{Derived in appendix~B of ref.~\cite{Oldengott_2015}.} $K(q,q',\mu)$ is defined as
\begin{align}
    K(q,q',\mu)
    &\equiv \frac{1}{16P^5}\mathrm{e}^{-(Q_-+P)/2}(Q_-^2-P^2)^2 \notag\\
    &\qquad \times \left[P^2(3P^2-2P-4) + Q_+^2(P^2+6P+12)\right], \label{eq:integration_kernel}
\end{align}
with the variables $Q_\pm \equiv q \pm q'$ and $P \equiv |\mathbf{q} - \mathbf{q}'|$, and where we write $\mu \equiv \cos\theta$ following \cref{fig:momentum_vectors}.
The dependence of $\mu$ is thus found within $P$, as this obeys the law of cosines:
\begin{gather}
  P^2 = q^2 + q'^2 - 2qq'\mu. \label{eq:law_of_cosines}
\end{gather}

\subsection{The Boltzmann hierarchy}
We now have all the ingredients for writing down the Boltzmann hierarchy that describes the Majorana neutrinos interacting via the exchange of a very massive scalar mediator particle.
This is done by decomposing \cref{eq:firstorder_massive_collision} in the Legendre expansion of $\Psi(\mathbf{k}, \mathbf{q})$ in $\mu$ and eliminating the azimuthal angle dependency between $\mathbf{k}$ and $\mathbf{q}$.
The decomposition is then completed by utilising properties of the spherical harmonics\footnote{Derived in appendix~C of ref.~\cite{Oldengott_2015}.}.

\begin{figure}[tb]
    \centering
    \includegraphics[width=\columnwidth]{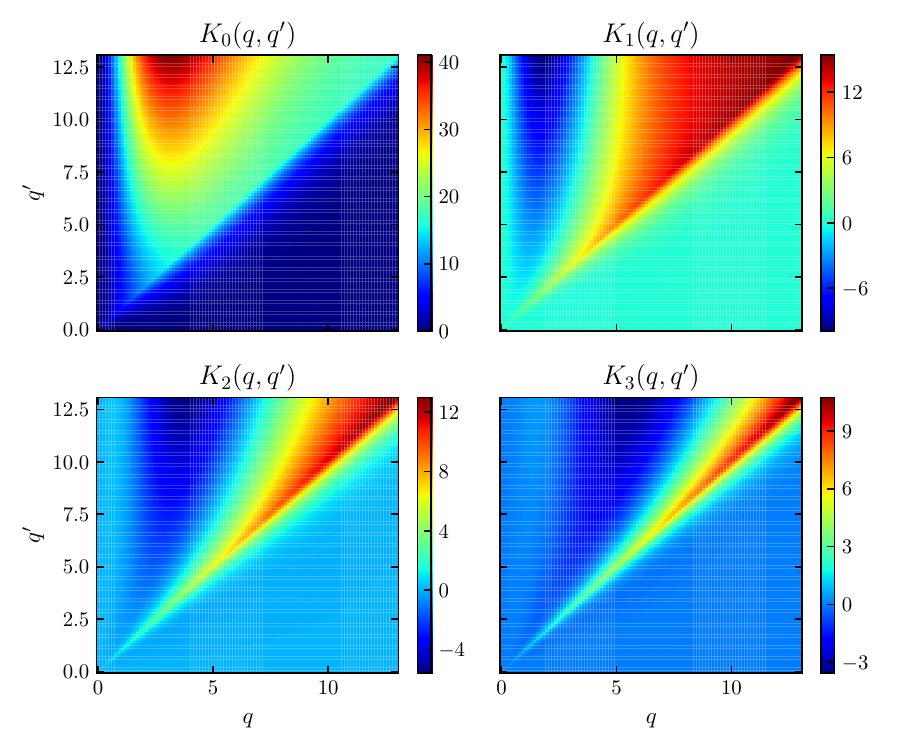}
    \caption{
      Visualisation of the collision integral kernels for multipoles $\ell=0,1,2,3$.
      As mentioned, the momentum variables $q$ and $q'$ are in units of the present-day neutrino temperature $T_{\nu,0}$.
    }
    \label{fig:integral_kernels}
\end{figure}

\noindent
From now on, we will absorb the present-day neutrino temperature $T_{\nu,0}$ in the comoving momentum $q$ to work with dimensionless quantities.
The hierarchy then takes the form
\begin{align}
  \dot{\Psi}_0(q)
  &= -k\Psi_1(q) + \frac{1}{6}\frac{\partial\ln \bar{f}}{\partial\ln q}\dot{h}
  - \frac{80}{3}\frac{\mathrm{N}T_{\nu,0}^5G_\mathrm{eff}^2}{a^4(2\pi)^3}q\Psi_0(q) \notag\\
  &\qquad + \frac{4\mathrm{N}T_{\mu,0}^5G_\mathrm{eff}^2}{a^4(2\pi)^3}\int\left[K_0(q,q')-\frac{10}{9}q^2q'^2\mathrm{e}^{-q}\right]\frac{q'\bar{f}(q')}{q\bar{f}(q)}\Psi_0(q')\,\mathrm{d}q' \label{eq:Boltzmannl0}\\
  \dot{\Psi}_1(q) &= -\frac{2}{3}k\Psi_2(q) + \frac{1}{3}k\Psi_0(q) - \frac{80}{3}\frac{\mathrm{N}T_{\nu,0}^5G_\mathrm{eff}^2}{a^4(2\pi)^3}q\Psi_1(q) \notag\\
  &\qquad + \frac{4\mathrm{N}T_{\nu,0}^5G_\mathrm{eff}^2}{a^4(2\pi)^3}\int\left[K_1(q,q')+\frac{5}{9}q^2q'^2\mathrm{e}^{-q}\right]\frac{q'\bar{f}(q')}{q\bar{f}(q)}\Psi_1(q')\,\mathrm{d}q' \label{eq:Boltzmannl1} \\
  \dot{\Psi}_2(q) &= -\frac{3}{5}k\Psi_3(q) + \frac{2}{5}k\Psi_1(q) - \frac{\partial\ln\bar{f}}{\partial\ln q}\left(\frac{2}{5}\dot{\eta}+\frac{1}{15}\dot{h}\right) - \frac{80}{3}\frac{\mathrm{N}T_{\nu,0}^5G_\mathrm{eff}^2}{a^4(2\pi)^3}q\Psi_2(q) \notag\\
  &\qquad + \frac{4\mathrm{N}T_{\nu,0}^5G_\mathrm{eff}^2}{a^4(2\pi)^3}\int\left[K_2(q,q')-\frac{1}{9}q^2q'^2\mathrm{e}^{-q}\right]\frac{q'\bar{f}(q')}{q\bar{f}(q)}\Psi_2(q')\,\mathrm{d}q' \label{eq:Boltzmannl2}\\
  \dot{\Psi}_{\ell>2}(q) &= \frac{k}{2\ell+1}\left[\ell\Psi_{\ell-1}(q) - (\ell+1)\Psi_{\ell+1}(q)\right] - \frac{80}{3}\frac{\mathrm{N}T_{\nu,0}^5G_\mathrm{eff}^2}{a^4(2\pi)^3}q\Psi_\ell(q) \notag\\
  &\qquad + \frac{4\mathrm{N}T_{\nu,0}^5G_\mathrm{eff}^2}{a^4(2\pi)^3}\int \frac{q'\bar{f}(q')}{q\bar{f}(q)}K_\ell(q,q')\Psi_\ell(q')\,\mathrm{d}q',\label{eq:Boltzmannl3}
\end{align}
where the integral kernels $K_\ell(q, q')$ are the $\ell$\textsuperscript{th} Legendre moments of $K(q, q', \mu)$ given by
\begin{gather}
  K_\ell(q, q') = \int_{-1}^1 K(q, q', \mu) P_\ell(\mu) \,\mathrm{d}\mu. \label{eq:ellth_integration_kernel}
\end{gather}
$K_\ell(q,q')$ thus encapsulates the dynamics of the collision at different multipoles and is shown for the first four multipoles on \cref{fig:integral_kernels}.
To make analytical progress on \cref{eq:ellth_integration_kernel}, ref.~\cite{Oldengott_2015} suggests that one does a change of integration variable from $\mu$ to $P$ via the law of cosines, \cref{eq:law_of_cosines}, such that
\begin{gather}
  K_\ell(q, q') = \frac{1}{qq'} \int_{|Q_-|}^{Q_+} K(q, q', P) P_\ell\Bigg(\frac{q^2+q'^2-P^2}{2qq'}\Bigg) P \,\mathrm{d}P.
\end{gather}
However, this integral turns out to be equally intractable with direct methods, so a different strategy is needed.



\section{From the Boltzmann hierarchy to the master integral}\label{sec:boltzmantomaster}
In this section, we outline how to derive the explicit collision kernel we wish to calculate for the effectively massless neutrinos scattering via the exchange of a very massive mediator.
\subsection{Momentum averaging of the hierarchy}
Ignoring the metric terms, the $\ell$\textsuperscript{th} moment expansion of the Boltzmann hierarchy takes the form
\begin{gather}
  \dot{\Psi}_\ell(q) = \frac{k}{2\ell+1}\left[\ell\Psi_{\ell-1}(q) - (\ell+1)\Psi_{\ell+1}(q)\right] + \mathcal{C}[\Psi_\ell](q), \label{eq:Psi_hierarchy}
\end{gather}
where $\mathcal{C}[\Psi_\ell](q)$ is the collision operator
\begin{gather}
  \mathcal{C}[\Psi_\ell](q) = \frac{\mathrm{N}T_{\nu,0}^5G_\mathrm{eff}^2}{a^4(2\pi)^3} \left[ 4\int \frac{q'\bar{f}(q')}{q\bar{f}(q)}\left[K_\ell(q,q') - S_\ell(q,q') \right]\Psi_\ell(q')\,\mathrm{d}q' - \frac{80}{3}q\Psi_\ell(q) \right]\,, \label{eq:collision_operator}
\end{gather}
and we defined $S_\ell(q,q')$ to match \cref{eq:Boltzmannl0,eq:Boltzmannl1,eq:Boltzmannl2,eq:Boltzmannl3}, i.e.
\begin{align}
S_\ell(q,q') &\equiv q^2 q'^2 e^{-q} \times \begin{cases} \frac{10}{9} & \ell=0, \\ -\frac{5}{9} & \ell=1, \\ \frac{1}{9} & \ell=2, \\ 0 & \ell>2. \end{cases}
\end{align}
Our goal from here on is to rewrite this in terms of the momentum-averaged moments $\mathcal{F}_\ell$, such that we can match it with equation (2.7) of ref.~\cite{Oldengott_2017}, thereby deriving a formula for the model-dependent $\alpha_\ell$ coefficients used in numerical implementations of this cosmological model in \class~\cite{Blas:2011rf}.

To do so, we will utilise eq. (46) of ref.~\cite{Ma_1995}, but with the $\ell$\textsuperscript{th} moment expansion instead of the full perturbation.
The first step is to multiply \cref{eq:Psi_hierarchy} with $q \bar{f}(q) q^2 \,\mathrm{d}q$ and integrate this over $q \geq 0$.
Since the background distribution is time-independent under the assumption of Maxwell-Boltzmann statistics, we may move the conformal time derivative operator outside the integral.
Then, we divide by the same weight $q\bar{f}(q)q^2\,\mathrm{d}q$ and integrate that over $q \geq 0$.
Ignoring the collision term $\mathcal{C}[\Psi_\ell](q)$ for now, we obtain
\begin{align}
  \frac{\int q \bar{f}(q) q^2 \dot{\Psi}_\ell(q) \,\mathrm{d}q}{\int q \bar{f}(q) q^2 \,\mathrm{d}q}
  &= \frac{k}{2\ell+1} \frac{\int q \bar{f}(q) q^2 \left[\ell\Psi_{\ell-1}(q) - (\ell+1)\Psi_{\ell+1}(q)\right] \,\mathrm{d}q}{\int q \bar{f}(q) q^2 \,\mathrm{d}q} \\
  \frac{\partial}{\partial\tau} \frac{\int q \bar{f}(q) q^2 \Psi_\ell(q) \,\mathrm{d}q}{\int q \bar{f}(q) q^2 \,\mathrm{d}q}
  &= \frac{k}{2\ell+1} \left[ \ell \frac{\int q \bar{f}(q) q^2 \Psi_{\ell-1}(q) \,\mathrm{d}q}{\int q \bar{f}(q) q^2 \,\mathrm{d}q} - (\ell+1)\frac{\int q \bar{f}(q) q^2 \Psi_{\ell+1}(q) \,\mathrm{d}q}{\int q \bar{f}(q) q^2 \,\mathrm{d}q} \right] \\
  \dot{\mathcal{F}}_\ell
  &= \frac{k}{2\ell+1} \left[ \ell\mathcal{F}_{\ell-1} - (\ell+1)\mathcal{F}_{\ell+1} \right]. \label{eq:averaged_collisionless_hierarchy}
\end{align}
Including the collision term is then as simple as adding the momentum-averaged expression of it to \cref{eq:averaged_collisionless_hierarchy}, i.e.
\begin{gather}
  \dot{\mathcal{F}}_\ell
  = \frac{k}{2\ell+1} \left[ \ell\mathcal{F}_{\ell-1} - (\ell+1)\mathcal{F}_{\ell+1} \right]
  + \frac{\int q^2 \,\mathrm{d}q\, q \bar{f}(q) \mathcal{C}[\Psi_\ell](q)}{\int q^2 \,\mathrm{d}q\, q\bar{f}(q)}. \label{eq:averaged_hierarchy}
\end{gather}
The weight integral in the denominator is trivial, since $\bar{f}(q)$ takes the Maxwell-Boltzmann form of \cref{eq:neutrino_background} (see eg. equation~(3.351.3) in \cite{Gradshteĭn_1994}):
\begin{gather}
  \int_0^\infty q^2 \,\mathrm{d}q\, q\bar{f}(q) = \mathrm{N} \int_0^\infty \mathrm{d}q\, q^3 \mathrm{e}^{-q} = 6\mathrm{N}.
\end{gather}
Recall now equation~(2.7) of ref.~\cite{Oldengott_2017}:
\begin{gather}
  \dot{\mathcal{F}}_\ell = \frac{k}{2\ell+1} \left[ \ell\mathcal{F}_{\ell-1} - (\ell+1)\mathcal{F}_{\ell+1} \right] + \alpha_\ell \dot{\kappa}_\nu \mathcal{F}_\ell,
\end{gather}
where $\dot{\kappa}_\nu$ is the rate of change of the neutrino opacity given by \cref{eq:neutrino_opacity}.
Clearly, this must match with \cref{eq:averaged_hierarchy}, or in other words,
\begin{gather}
  \alpha_\ell \dot{\kappa}_\nu \mathcal{F}_\ell = \frac{1}{6\mathrm{N}} \int q \bar{f}(q) q^2 \mathcal{C}[\Psi_\ell](q) \,\mathrm{d}q \label{eq:hierarchies_matched}.
\end{gather}
\subsection{Separable ansatz and the coefficient $\alpha_\ell$}
To proceed, we will enforce the ``separable ansatz'', which implies that all the momentum dependence factors into the background distribution $\bar{f}(q)$, while the momentum-averaged moments $\mathcal{F}_\ell$ solely encapsulates the amplitudes.
The separable ansatz is well described in \cite{Cyr_Racine_2014} and allows us to write the $\ell$\textsuperscript{th} moment expansion of the Boltzmann hierarchy as
\begin{gather}
  \Psi_\ell(q) \sim -\frac{1}{4} \frac{\mathrm{d}\ln\bar{f}(q)}{\mathrm{d}\ln q} \mathcal{F}_\ell. \label{eq:separable_ansatz}
\end{gather}
As $\mathrm{d}\ln q = \mathrm{d}q/q$, the logarithmic derivative evaluates to
\begin{gather}
  \frac{\mathrm{d}\ln\bar{f}(q)}{\mathrm{d}\ln q}
  = q \frac{\mathrm{d}\ln\bar{f}(q)}{\mathrm{d}q}
  = q \frac{\mathrm{d}}{\mathrm{d}q}(\ln\mathrm{N} - q)
  = -q,
\end{gather}
and hence
\begin{gather}
  \Psi_\ell(q) \sim \frac{1}{4} q \mathcal{F}_\ell.
\end{gather}
Inserting this in \cref{eq:collision_operator}, we get an expression for the collision operator under the separable ansatz,
\begin{gather}
  \mathcal{C}[\Psi_\ell](q)
  = \frac{\mathrm{N}T_{\nu,0}^5G_\mathrm{eff}^2}{a^4(2\pi)^3}\mathcal{F}_\ell \left( \int \frac{q'^2\bar{f}(q')}{q\bar{f}(q)} \left[K_\ell(q,q') - S_\ell(q,q') \right]\,\mathrm{d}q' - \frac{20}{3}q^2 \right) \label{eq:collision_operator_separable_ansatz}.
\end{gather}
Now we insert \cref{eq:neutrino_opacity,eq:neutrino_background,eq:ellth_integration_kernel,eq:collision_operator_separable_ansatz} in \cref{eq:hierarchies_matched} to arrive at an explicit formula for the $\alpha_\ell$ coefficients:
\begin{align}
  \alpha_\ell
  &= \frac{1}{6\mathrm{N}} \frac{-1}{aG_\mathrm{eff}^2T_\nu^5\mathcal{F}_\ell} \int q \bar{f}(q) q^2 \frac{\mathrm{N}T_{\nu,0}^5G_\mathrm{eff}^2}{a^4(2\pi)^3}\mathcal{F}_\ell \left( \int \frac{q'^2\bar{f}(q')}{q\bar{f}(q)} \left[K_\ell(q,q') - S_\ell(q,q') \right] \,\mathrm{d}q' - \frac{20}{3}q^2 \right) \mathrm{d}q \,,\nonumber\\
  &= \frac{1}{6(2\pi)^3} \int \left( \frac{20}{3} q^5 \bar{f}(q) - \int  q^2 q'^2 \bar{f}(q') \left[K_\ell(q,q') - S_\ell(q,q') \right] \,\mathrm{d}q' \right) \mathrm{d}q \,,\nonumber \\
  &= \frac{\mathrm{N}}{6(2\pi)^3} \int \left( \frac{20}{3} q^5 \mathrm{e}^{-q} - \int q^2 q'^2 \mathrm{e}^{-q'} K(q, q', \mu) P_\ell(\mu) \,\mathrm{d}\mu \,\mathrm{d}q'  + \int  q^2 q'^2  \mathrm{e}^{-q'}  S_\ell(q,q') \,\mathrm{d}q' \right) \mathrm{d}q \,,\nonumber \\
   &= \frac{\mathrm{N}}{6(2\pi)^3} (800 - I_\ell + s_\ell), \label{eq:alpha_ell_by_I_ell}
\end{align}
where the first term is evaluated via, once again, equation~(3.351.3) in \cite{Gradshteĭn_1994}, and $I_\ell$ encapsulates everything that still involves integrals.
Similar to the first term, the $s_\ell$ integrals can be trivially evaluated.
They are equal to
\begin{align}
s_0 &=  640\,, \qquad s_1 = -320\,, \qquad s_2 = 64 \,, \qquad s_{\ell>2} = 0. \label{eq:sellvalues}
\end{align}

\subsection{The master integral}\label{eq:masterintegral}
The integral $I_\ell$ is implicitly defined in \cref{eq:alpha_ell_by_I_ell}. Inserting the integration kernel \cref{eq:integration_kernel} yields
\begin{align}
  I_\ell
  &\equiv \int q^2 q'^2 \mathrm{e}^{-q'} K(q, q', \mu) P_\ell(\mu) \,\mathrm{d}\mu \,\mathrm{d}q' \,\mathrm{d}q \label{eq:I_ell_definition}\\
  &= \int q^2 q'^2 \mathrm{e}^{-q'} \frac{1}{16P^5}\mathrm{e}^{-(Q_-+P)/2}(Q_-^2-P^2)^2 \notag\\
  &\qquad \times \left[P^2(3P^2-2P-4) + Q_+^2(P^2+6P+12)\right] \,\mathrm{d}\mu \,\mathrm{d}q' \,\mathrm{d}q \notag\\
  &= \int q^2 q'^2 \mathrm{e}^{-q/2} \mathrm{e}^{-q'/2} \mathrm{e}^{-P/2} \frac{1}{16P^5} (Q_-^2-P^2)^2 \notag\\
  &\qquad \times \left[ P^2(3P^2-2P-4) + Q_+^2(P^2+6P+12) \right] P_\ell(\mu) \,\mathrm{d}\mu \,\mathrm{d}q' \,\mathrm{d}q.
\end{align}
By rewriting $(Q_-^2 - P^2)^2$ via the law of cosines, \cref{eq:law_of_cosines},
\begin{align}
  (Q_-^2 - P^2)^2 &= (q^2 + q'^2 - 2qq' - q^2 - q'^2 + 2qq'\mu)^2 \\
  &= (2qq'\mu - 2qq')^2 \\
  &= [2qq'(\mu - 1)]^2 \\
  &= 4q^2 q'^2 (\mu - 1)^2,
\end{align}
we can rewrite $I_\ell$ in its explicit, $(q,q')$-symmetric form, namely
\begin{align}
  I_\ell &= \int_0^\infty \int_0^\infty \int_{-1}^1 4q^4 q'^4 \mathrm{e}^{-q/2} \mathrm{e}^{-q'/2} \mathrm{e}^{-P/2} (\mu - 1)^2 \frac{1}{16P^5} \notag\\
  &\qquad \times \left[ P^2(3P^2-2P-4) + Q_+^2(P^2+6P+12) \right] P_\ell(\mu) \,\mathrm{d}\mu \,\mathrm{d}q' \,\mathrm{d}q. \label{eq:integral_symmetric}
\end{align}
The first few exact values of this integral are
\begin{align}
  I_0 &= 1440, \notag\\
  I_1 &= 480, \notag\\
  I_2 &= -2273760+230400\pi^2, \notag\\
  I_3 &= 181916640-18432000\pi^2, \notag\\
  I_4 &= -4968595680+503424000\pi^2, \notag\\
  I_5 &= 75768242400-7676928000\pi^2, \notag\\
  I_6 &= -781638559200+79196544000\pi^2, \notag\\
  I_7 &= 6047977099872-612788198400\pi^2, \notag\\
  I_8 &= -37440097002720+3793474944000\pi^2\,,
  \label{eq:first_integral_values}
\end{align}
which seems to suggest that the integral will always be the difference between an integer and another integer multiplied by $\pi^2$. However, starting from $\ell=9$, the first term becomes a rational number. The second term on the other hand seems to remain an integer; at least it is still an integer for $\ell=50,000$. Direct symbolic integration of \cref{eq:integral_symmetric} in \textsc{Mathematica} is doable up to roughly $\ell \lesssim 17$, but it is a delicate process since many divergences must cancel. It is also worth noting that $I_0$ and $I_1$, together with $s_1$ and $s_2$ from \cref{eq:sellvalues} makes $\alpha_0 = \alpha_1 = 0$ when inserted into \cref{eq:alpha_ell_by_I_ell} as expected from energy and momentum conservation.


\section{Analytic reduction of the kernel} \label{sec:recursion}
The remaining task is the evaluation of the dimensionless integral $I_\ell$ in \cref{eq:integral_symmetric}. The key insight is that the full integral can be reduced to a finite sum of simpler momentum integrals whose recurrence relations are stable and easy to implement.

The strategy has three ingredients.  First, the complicated integration kernel is written as $\mu$-derivatives of the base-kernel $\frac{\mathrm{e}^{-P/2}}{P}$ which turns out to be analytically tractable. Second, integration by parts moves these derivatives onto Legendre polynomials, producing three finite angular mixing operators.  Third, the momentum integrals generated by these operators are reduced to one base family, $\mathcal{I}^{4,4}_k$, by recurrence relations.

\subsection{The base momentum integrals}
We use the symmetry under $q\leftrightarrow q'$ to work on the triangular domain $0<q'<q<\infty$ and multiply by two.  On this domain the addition theorem for modified spherical Bessel functions, \href{http://dlmf.nist.gov/10.60.E3}{DLMF eq. (10.60.3)}, gives
\begin{gather}
  \frac{\mathrm{e}^{-P/2}}{P}
  = \frac{1}{\pi}\sum_{k=0}^{\infty}(2k+1)
  i_k\!\left(\frac{q'}{2}\right) k_k\!\left(\frac{q}{2}\right)P_k(\mu),
  \label{eq:base_bessel_expansion}
\end{gather}
where $i_k$ and $k_k$ are modified spherical Bessel functions. In this paper we use $i_k \equiv i_k^{(1)}$, i.e. $i_k$ is always of the first kind.  After the rescaling $q\mapsto 2q$ and $q'\mapsto 2q'$, the momentum objects needed below are
\begin{gather}
  \mathcal{I}^{m,n}_k
  \equiv \frac{2^{m+n+3}}{\pi}
  \int_0^\infty \! \mathrm{d}q \int_0^q \! \mathrm{d}q'\,
  q^m q'^n \mathrm{e}^{-q-q'} i_k(q')k_k(q).
  \label{eq:base_integral_definition}
\end{gather}
The factor $2^{m+n+3}/\pi$ is included so that the final formula below contains only rational angular weights.
The leading member, $\mathcal{I}^{4,4}_k$, has the closed form given in appendix~\ref{app:base_integral}.  For computation it is more convenient to use its recurrence relation.  The seed values are
\begin{align}
  \mathcal{I}^{4,4}_0 &= 2484, \qquad
  \mathcal{I}^{4,4}_1 = 2060, \qquad
  \mathcal{I}^{4,4}_2 = 1620, \qquad
  \mathcal{I}^{4,4}_3 = 1260, \notag\\
  \mathcal{I}^{4,4}_4 &= -4641676 + 470400\pi^2,
  \label{eq:I44_seed_values}
\end{align}
while, for $k\geq 5$,
\begin{gather}
  \mathcal{I}^{4,4}_k
  = -\frac{(k+4)^2}{(k-4)^2}\mathcal{I}^{4,4}_{k-1}
  + \frac{80640}{(k-4)^2}.
  \label{eq:I44_recurrence}
\end{gather}
Since the recurrence coefficients are rational, the decomposition
\begin{gather}
  \mathcal{I}^{4,4}_k = r_k + s_k\pi^2,
  \qquad r_k,s_k\in\mathbb{Q},
  \label{eq:I44_split}
\end{gather}
is preserved exactly.
The reduction of the full kernel also produces $\mathcal{I}^{4,2}_k+\mathcal{I}^{2,4}_k$ and $\mathcal{I}^{3,3}_k$.  These do not need an independent treatment, since they are related to $\mathcal{I}^{4,4}_k$ by
\begin{align}
  \mathcal{I}^{4,2}_k+\mathcal{I}^{2,4}_k &= a_k\mathcal{I}^{4,4}_k+b_k,
  \label{eq:I42I24_reduction}\\
  \mathcal{I}^{3,3}_k &= c_k\mathcal{I}^{4,4}_k+d_k,
  \label{eq:I33_reduction}
\end{align}
where
\begin{align}
  a_k &=
  \begin{cases}
    0, & k<4,\\[2pt]
    \displaystyle\frac{24}{(k-3)(k-2)(k+3)(k+4)}, & k\geq4,
  \end{cases}
  \label{eq:a_coefficient}\\
  b_k &=
  \begin{cases}
    168, & k=0,\\
    120, & k=1,\\
    84, & k=2,\\
    60, & k=3,\\[2pt]
    \displaystyle\frac{1440(k^2+k-33)}{(k-3)(k-2)(k+3)(k+4)}, & k\geq4,
  \end{cases}
  \label{eq:b_coefficient}\\
  c_k &=
  \begin{cases}
    0, & k<4,\\[2pt]
    \displaystyle\frac{16}{(k-3)^2(k+4)^2}, & k\geq4,
  \end{cases}
  \label{eq:c_coefficient}\\
  d_k &=
  \begin{cases}
    76, & k=0,\\
    56, & k=1,\\
    40, & k=2,\\
    23716-2400\pi^2, & k=3,\\[2pt]
    \displaystyle\frac{720(k^2+k-40)}{(k^2+k-12)^2}, & k\geq4.
  \end{cases}
  \label{eq:d_coefficient}
\end{align}
The derivation of \eqref{eq:I44_recurrence}, \eqref{eq:I42I24_reduction}, and \eqref{eq:I33_reduction} is summarised in appendix~\ref{app:recurrences}.  For $k<4$ the $b_k$ and $d_k$ entries were found from direct evaluations of the corresponding integrals. This is necessary since the general recurrences are not well-defined for small $k$. 

\subsection{Removing the high inverse powers of $P$}
The apparent obstacle in \cref{eq:integral_symmetric} is the factor $P^{-5}$.  This can be eliminated before doing the angular integral. From the law of cosines we find,
\begin{gather}
  \frac{\mathrm{d}}{\mathrm{d}\mu}
  = -\frac{qq'}{P}\frac{\mathrm{d}}{\mathrm{d}P},
  \qquad
  \frac{\mathrm{d}^2}{\mathrm{d}\mu^2}
  = q^2q'^2\left(\frac{1}{P}\frac{\mathrm{d}}{\mathrm{d}P}\right)^2.
  \label{eq:mu_derivative_operator}
\end{gather}
Acting on the two simple functions $\mathrm{e}^{-P/2}/P$ and $P\mathrm{e}^{-P/2}$ gives
\begin{align}
  \frac{\mathrm{d}^2}{\mathrm{d}\mu^2}\left(\frac{\mathrm{e}^{-P/2}}{P}\right)
  &= 4q^2q'^2\frac{\mathrm{e}^{-P/2}}{16P^5}\left(P^2+6P+12\right),
  \label{eq:derivative_base_kernel}\\
  \frac{\mathrm{d}^2}{\mathrm{d}\mu^2}\left(P\mathrm{e}^{-P/2}\right)
  &= 4q^2q'^2\frac{\mathrm{e}^{-P/2}}{16P^5}P^2\left(3P^2-2P-4\right)
  - q^2q'^2\frac{\mathrm{e}^{-P/2}}{2P}.
  \label{eq:derivative_P_kernel}
\end{align}
Using these identities in the kernel and replacing $P\mathrm{e}^{-P/2}=P^2\mathrm{e}^{-P/2}/P$ gives the exact decomposition
\begin{align}
  \mathrm{e}^{Q_-/2}K(q,q',\mu)
  &= 2(1-\mu)^2\frac{\mathrm{d}^2}{\mathrm{d}\mu^2}
  \left(\left[q^2+q'^2+qq'(1-\mu)\right]\frac{\mathrm{e}^{-P/2}}{P}\right) \notag\\
  &\qquad + q^2q'^2(1-\mu)^2\frac{\mathrm{e}^{-P/2}}{2P}.
  \label{eq:kernel_derivative_decomposition}
\end{align}
Thus every remaining angular integral involves only the base kernel $\mathrm{e}^{-P/2}/P$.

It is convenient to define the angular function $u_\ell$,
\begin{gather}
  u_\ell(\mu)\equiv (1-\mu)^2P_\ell(\mu).
  \label{eq:u_ell_definition}
\end{gather}
Integrating the first term of \cref{eq:kernel_derivative_decomposition} by parts twice moves the $\mu$-derivatives onto $u_\ell$.
The endpoint at $\mu=1$ vanishes because both $u_\ell$ and $u_\ell'$ vanish there.
At $\mu=-1$, where $P=q+q'$, the remaining boundary term can be evaluated analytically and is simply

\begin{gather}
  B_\ell = -48(-1)^\ell\left[7+2\ell(\ell+1)\right] \,.
  \label{eq:boundary_integrated}
\end{gather}
After the triangular-domain factor of two is included, the physical bulk angular operators multiplying $\mathcal{I}^{4,4}_k$, $\mathcal{I}^{4,2}_k+\mathcal{I}^{2,4}_k$, and $\mathcal{I}^{3,3}_k$ are
\begin{gather}
  \mathcal{I}^{4,4}_k = u_\ell(\mu),\qquad
  \mathcal{I}^{4,2}_k+\mathcal{I}^{2,4}_k = 4u_\ell''(\mu),\qquad
  \mathcal{I}^{3,3}_k = 4(1-\mu)u_\ell''(\mu).
  \label{eq:physical_angular_operators}
\end{gather}
The bulk terms are controlled by three angular operators,
\begin{align}
  O^{(1)}_\ell(\mu) &= u_\ell(\mu),
  \label{eq:operator_1}\\
  O^{(2)}_\ell(\mu) &= 4u_\ell''(\mu),
  \label{eq:operator_2}\\
  O^{(3)}_\ell(\mu) &= 4(1-\mu)u_\ell''(\mu).
  \label{eq:operator_3}
\end{align}
We define the weights $W^{(i)}_{\ell k}$ as the coefficients of their Legendre expansions,
\begin{gather}
  O^{(i)}_\ell(\mu)=\sum_{k\geq0}W^{(i)}_{\ell k}P_k(\mu),
  \qquad
  W^{(i)}_{\ell k}=\frac{2k+1}{2}\int_{-1}^{1}O^{(i)}_\ell(\mu)P_k(\mu)\,\mathrm{d}\mu.
  \label{eq:W_definition}
\end{gather}
The explicit forms are
\begin{align}
  W^{(1)}_{\ell k} &= \frac{2k+1}{2}
  \begin{cases}
    \displaystyle \frac{2(\ell+1)(\ell+2)}{(2\ell+1)(2\ell+3)(2\ell+5)}, & k=\ell+2,\\[6pt]
    \displaystyle \frac{-4(\ell+1)}{(2\ell+1)(2\ell+3)}, & k=\ell+1,\\[6pt]
    \displaystyle \frac{4(3\ell^2+3\ell-2)}{(2\ell-1)(2\ell+1)(2\ell+3)}, & k=\ell,\\[6pt]
    \displaystyle \frac{-4\ell}{(2\ell-1)(2\ell+1)}, & k=\ell-1,\\[6pt]
    \displaystyle \frac{2\ell(\ell-1)}{(2\ell-3)(2\ell-1)(2\ell+1)}, & k=\ell-2,\\[6pt]
    0, & \mathrm{otherwise},
  \end{cases}
  \label{eq:W1_cases}\\
  W^{(2)}_{\ell k} &= \frac{2k+1}{2}
  \begin{cases}
    0, & k>\ell,\\[6pt]
    \displaystyle \frac{8(\ell+1)(\ell+2)}{2\ell+1}, & k=\ell,\\[6pt]
    \displaystyle 8(-1)^{\ell+k}\left[\ell(\ell+1)-k(k+1)+2\right], & k<\ell,
  \end{cases}
  \label{eq:W2_cases}\\
  W^{(3)}_{\ell k} &= \frac{2k+1}{2}
  \begin{cases}
    0, & k>\ell+1,\\[6pt]
    \displaystyle \frac{-8(\ell+1)^2(\ell+2)}{(2\ell+1)(2\ell+3)}, & k=\ell+1,\\[6pt]
    \displaystyle \frac{8(\ell+1)(3\ell+2)}{2\ell+1}, & k=\ell,\\[6pt]
    \displaystyle \frac{-8(17\ell^3+7\ell^2-4\ell-2)}{4\ell^2-1}, & k=\ell-1,\\[6pt]
    \displaystyle 16(-1)^{\ell+k}\left[\ell(\ell+1)-k(k+1)+1\right], & k<\ell-1.
  \end{cases}
  \label{eq:W3_cases}
\end{align}
The lower limit in \eqref{eq:W2_cases} and \eqref{eq:W3_cases} is genuinely $k=0$: the derivatives of $P_\ell$ generate all Legendre modes, not just a narrow band around $\ell$.  The derivation of \cref{eq:W1_cases,eq:W2_cases,eq:W3_cases} is given in appendix~\ref{app:angular_weights}.

\subsection{Exact expression for the integral}
With the definitions above, the full integral becomes
\begin{align}
  I_\ell
  &= B_\ell
  + \sum_{k=0}^{\ell+2}\bigg[
    W^{(1)}_{\ell k}\mathcal{I}^{4,4}_k
    + W^{(2)}_{\ell k}\left(\mathcal{I}^{4,2}_k+\mathcal{I}^{2,4}_k\right)
    + W^{(3)}_{\ell k}\mathcal{I}^{3,3}_k
  \bigg].
  \label{eq:exact_integral_unreduced}
\end{align}
Using \eqref{eq:I42I24_reduction} and \eqref{eq:I33_reduction}, this becomes the implementation formula
\begin{align}
  I_\ell
  &= B_\ell
  + \sum_{k=0}^{\ell+2}
  \left(W^{(1)}_{\ell k}+a_kW^{(2)}_{\ell k}+c_kW^{(3)}_{\ell k}\right)\mathcal{I}^{4,4}_k 
  + \sum_{k=0}^{\ell+2}\left(b_kW^{(2)}_{\ell k}+d_kW^{(3)}_{\ell k}\right).
  \label{eq:exact_integral_final}
\end{align}
All quantities entering \cref{eq:exact_integral_final} are rational except for the explicit $\pi^2$ pieces in $\mathcal{I}^{4,4}_k$ and $d_3$.  This means that we can compute the full result as two exact rational sums, one multiplying $1$ and one multiplying $\pi^2$.

Finally, the collision coefficient used in the momentum-averaged hierarchy follows from \cref{eq:alpha_ell_by_I_ell},
\begin{gather}
  \alpha_\ell = \frac{\mathrm{N}}{6(2\pi)^3}\left(800-I_\ell + s_\ell \right),
  \label{eq:alpha_final}
\end{gather}
with $s_\ell$ given by \cref{eq:sellvalues}. Having carried out the full integral analytically we are now in a position to compare the exact values of $\alpha_\ell$ with the effective values used in previous treatments \cite{Oldengott_2017}.  This comparison is shown in \cref{table:alpha_values} for values for $\alpha_\ell$ up to $\ell = 10$.

\begin{table}[tb]
    \begin{center} 
        \begin{tabular}{c c c c} 
            \hline
            $\ell$ & $\alpha_{\ell, {\rm eff}}$ \cite{Oldengott_2017} & $\alpha_{\ell, {\rm exact}}$ & $\alpha_{\ell, {\rm exact, rounded}}$  \\
            \hline
            $2$ &  $0.40$ & $\frac{7 \pi}{60}(3949-400 \pi^2)$ & $0.42452$\\
$3$ &  $0.43$ & $\frac{7 \pi}{108}(-568487+57600 \pi^2)$ & $0.45072$\\
$4$ &   $0.46$ & $\frac{28 \pi}{27}(970429-98325 \pi^2)$ & $0.47977$\\
$5$ & $0.47$  & $\frac{35 \pi}{108}(-47355151+4798080 \pi^2)$ & $0.49356$\\
$6$ & $0.48$  & $\frac{175 \pi}{27}(24426205-2474892 \pi^2)$ & $0.50049$\\
$7$ & $0.48$  & $\frac{7 \pi}{540}(-94499642173+9574815600 \pi^2)$ & $0.50413$\\
$8$ &  $0.48$ & $\frac{28 \pi}{27}(7312518946-740913075 \pi^2)$ & $0.50612$\\
$9$ & $0.48$  & $\frac{\pi}{756}(-29673467398493+3006550839600 \pi^2)$ & $0.50725$\\
$10$ & $0.48$  & $\frac{\pi}{756}(132564618268039-13431604032000 \pi^2)$ & $0.50791$\\
            \hline						
        \end{tabular}
    \end{center}
    \caption{Values of $\alpha$ for different $\ell$. Since $I_\ell \sim \ell^{-6}$, the asymptotic value is $\alpha_\ell \rightarrow \frac{35\pi}{216} \simeq 0.50905$. At five significant figures, this asymptotic value is achieved for $\ell \gtrsim 26$. $\alpha_0$ and $\alpha_1$ vanish due to energy and momentum conservation as mentioned at the end of section \cref{eq:masterintegral}.}
    \label{table:alpha_values}
\end{table}

\section{Observables - numerical results}\label{sec:observables}
We do not expect to see any observable effects from the small changes in $\alpha_\ell$ from using the exact values compared to the approximate values from equation 2.9 in ref.~\cite{Oldengott_2017}. Nevertheless, we have implemented the exact $\alpha_\ell$-values from \cref{eq:alpha_ell_by_I_ell} computed using the integral approximation \cref{eq:rational_fit} in \class{} to demonstrate this. We are computing this using the massive interacting neutrino hierarchy but with a mass $m_\nu 10^{-8}\text{eV}$. We also use a temperature $T_{\nu,0} = \left(\frac{4}{11}\right)^{\frac{1}{3}} T_{\gamma,0}$ and $\text{deg}_\nu = 3.04$, and we turned off the fluid approximation for this test.

\begin{figure}[tb]
    \centering
    \includegraphics[width=\columnwidth]{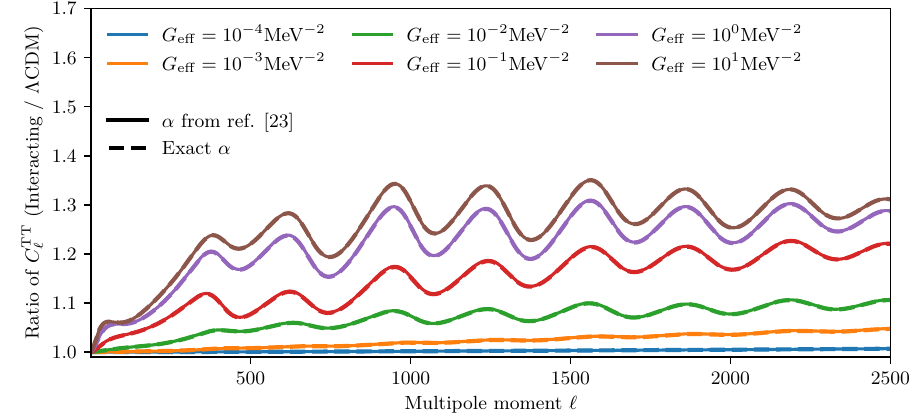}
    \caption{Ratio of angular temperature power spectra for three interacting massless neutrinos compared to $\Lambda \text{CDM}$. Solid lines use the values from ref.~\cite{Oldengott_2017} while the dashed line use the exact $\alpha_\ell$ from this paper. The differences are non-observable, as expected.}
    \label{fig:ClTTratio}
\end{figure}

We first plot the ratio of the angular temperature power spectrum, $C_\ell^\text{TT}$ between the interacting case and the $\Lambda$CDM-model, in this case realised by putting the coupling constant to $G_\text{eff} = 10^{-15}$. We see the usual enhancement of up to $\sim 30\%$ from removing anisotropic stress. However, the few percent differences in $\alpha_\ell$ do not lead to visually distinct lines. We may instead try to plot the relative effect on the $C_\ell^\text{TT}$ spectrum directly as we shall do next.

\begin{figure}[tb]
    \centering
    \includegraphics[width=\columnwidth]{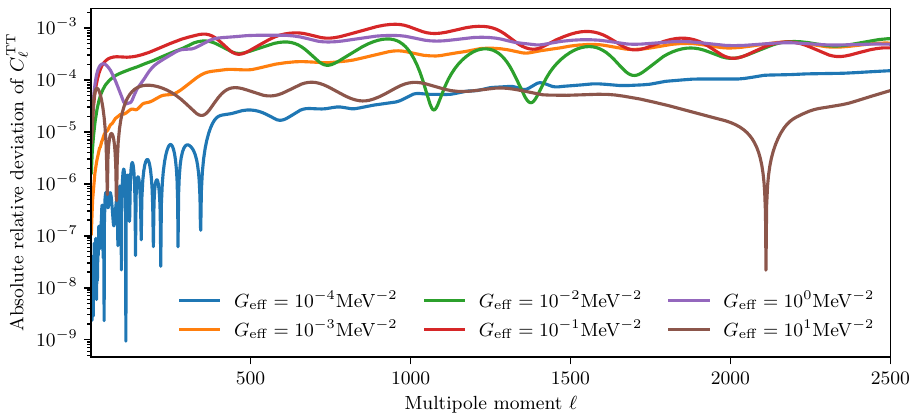}
    \caption{Relative deviations of the angular temperature power spectra when using $\alpha_\ell$ from ref.~\cite{Oldengott_2017} compared to the exact values for different coupling strengths.}
    \label{fig:ClTTratioInteracting}
\end{figure}

\Cref{fig:ClTTratioInteracting} shows the relative effect on $C_\ell^\text{TT}$ from changing $\alpha_\ell$ from the numerical values in ref.~\cite{Oldengott_2017} to the exact values. \class{} is not expected to be precise below $\sim 10^{-3}$, so we must be careful when interpreting these differences. However, for very small coupling strengths and for very large coupling strengths, we expect $\alpha_\ell$ to be largely irrelevant, since the system is either fully free-streaming or fully fluid-like. Both these curves are at the $10^{-4}$-$10^{-5}$ level, so we can cautiously interpret this as an optimistic noise-floor. That allows us to quantify the effect of correcting the $\alpha_\ell$ coefficients to be at most one per mille relative change in $C_\ell^\text{TT}$. This is far from observable of course, not to mention that it will be almost fully correlated with the effective coupling strength $G_\text{eff}$.

\begin{figure}[tb]
    \centering
    \includegraphics[width=\columnwidth]{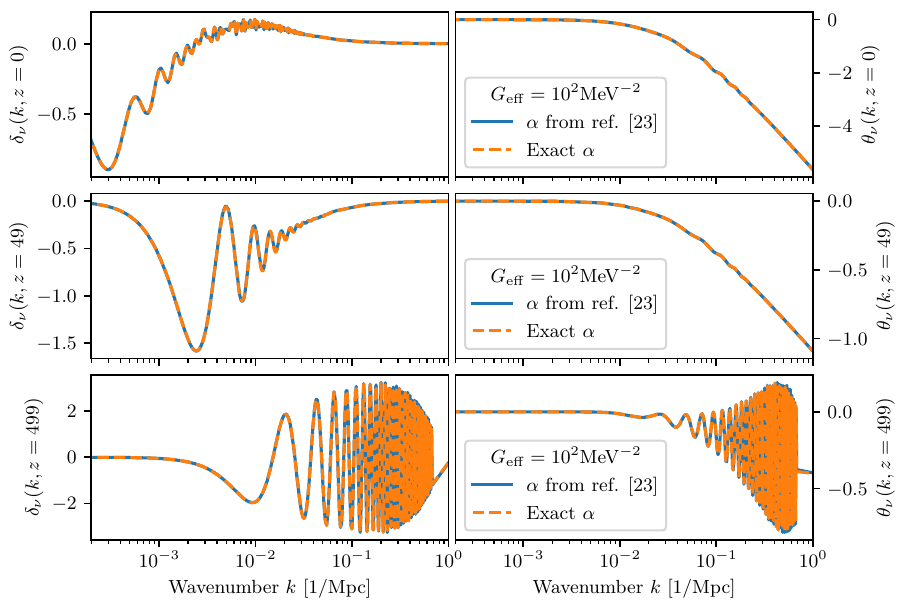}
    \caption{Density (left panels) and velocity transfer functions (right panels) with and without the exact $\alpha_\ell$ for redshifts $z=0, 49, 499$. The coupling strength was chosen to be $G_\text{eff} = 10^2 \text{MeV}^{-2}$ to push decoupling to later times.}
    \label{fig:transferfunctions}
\end{figure}

For completeness, we also compute the density and velocity transfer functions in the comoving synchronous gauge for redshifts $z=0, 49, 499$, again under both prescriptions for $\alpha_\ell$. There is again no visual difference between the two which is not unexpected.

\section{Discussion and conclusion}\label{sec:discussion}

Using the relaxation time approximation to the full collision term in the Boltzmann equation for massless (or very light) neutrinos interacting via the exchange of a very massive scalar or pseudo-scalar, we have managed to analytically integrate the collision term to provide an exact analytic expression for the $\alpha_\ell$ coefficients defined in \cite{Oldengott_2017}.

The reduction above replaces an unstable direct integration of the collision kernel by the finite algebraic expression in \cref{eq:exact_integral_final}.  The important point is that the angular and momentum problems are separated.  The angular dependence is carried by the three weight families $W^{(1)}_{\ell k}$, $W^{(2)}_{\ell k}$, and $W^{(3)}_{\ell k}$, while the momentum dependence is reduced to $\mathcal{I}^{4,4}_k$ and rational corrections.  This is the notation used in the implementation.


The practical implementation details, including the exact rational evaluation, the Chudnovsky conversion for $\pi^2$, and the asymptotically constrained fit, are collected in appendix~\ref{app:implementation_details}. It is possible that our analytical solutions could be useful outside of cosmology (or even Physics), although we did not find any clear candidate except from maybe ref.~\cite{Seth2012}.

Using the exact integration only leads to modest changes to the numerical coefficients used in codes such as \class{} or \camb, which in the end translate to only minute changes to actual neutrino observables such as $\delta_\nu$, $\theta_\nu$, and $\sigma_\nu$. A similar set of $\alpha_\ell$-coefficients has also been derived for inverse neutrino decays~\cite{Chen:2022idm}. In this case, the $\alpha_\ell$-coefficients are much larger due to the small angle involved in the inverse scattering process, and it would be interesting to see to what extent a fully analytical treatment could be carried out in this model. 


\section*{Reproducibility}
A self-contained reference implementation is available in the public companion notebook, \integralnotebook. It evaluates \cref{eq:exact_integral_final} using exact rational arithmetic, optionally with \texttt{gmpy2} for speed and Python's built-in \texttt{fractions.Fraction} as a fallback.  The implementation then converts the exact expression $R_\ell+S_\ell\pi^2$ to double precision using \eqref{eq:chudnovsky_sum} and \eqref{eq:pi2_chudnovsky}.

\section*{Acknowledgements}
We thank Yvonne Wong and Isabel Oldengott for feedback on our manuscript and Tobias Bach Johannessen for implementing the corrected $\alpha_\ell$ in \class{}. We acknowledge computing resources from the Centre for Scientific Computing Aarhus (CSCAA).

\appendix

\section{Closed form for the base integral}\label{app:base_integral}
All prefactors in this appendix use the normalisation of \cref{eq:base_integral_definition}.  If the inner momentum integral is done first, the base integral can be written as
\begin{align}
  \mathcal{I}^{4,4}_\ell
  &=
  \frac{6\ell!(\ell+4)!}{\pi\sqrt{\pi}}
  \int_0^\infty\mathrm{d}q\,
  \mathrm{e}^{-q/2}q^{\ell+17/2}
  K_{\ell+1/2}\left(\frac{q}{2}\right)
  {}_2\tilde{F}_2\left(\ell+1,\ell+5;\ell+6,2\ell+2;-q\right),
  \label{eq:I44_after_inner_integral}
\end{align}
where ${}_p\tilde{F}_q$ denotes the regularised hypergeometric function.  Since the Bessel index is half-integer,
\begin{gather}
  K_{\ell+1/2}\left(\frac{q}{2}\right)
  =
  \sqrt{\frac{\pi}{q}}\mathrm{e}^{-q/2}
  \sum_{r=0}^{\ell}\frac{(\ell+r)!}{r!(\ell-r)!}\frac{1}{q^r}.
  \label{eq:K_half_integer_appendix}
\end{gather}
Term-by-term integration then gives the finite sum
\begin{align}
  \mathcal{I}^{4,4}_\ell
  &=
  \sum_{r=0}^{\ell}
  \frac{2\ell!(\ell+4)!(\ell+r)!(\ell-r+8)!}{r!(\ell-r)!}
  {}_3\tilde{F}_2\left(\ell+1,\ell+5,\ell-r+9;\ell+6,2\ell+2;-1\right).
  \label{eq:I44_finite_sum}
\end{align}
Summing over $r$ collapses the finite sum to a single hypergeometric function,
\begin{align}
  \mathcal{I}^{4,4}_\ell
  &=
  \frac{
  80640\,\Gamma(\ell+1)\Gamma(\ell+5)\Gamma(2\ell+10)}
  {\Gamma(2\ell+2)\Gamma(\ell+6)\Gamma(\ell+10)}
  \notag\\
  &\quad\times
  {}_4F_3\left(9,\ell+1,\ell+5,2\ell+10;\ell+6,\ell+10,2\ell+2;-1\right).
  \label{eq:I44_4F3}
\end{align}
The difficult step is the reduction of this ${}_4F_3$.  We performed it with \texttt{pfq.m}, the ${}_pF_q$ component of the HYPERDIRE package~\cite{Bytev:2011ks}.  The reduction leaves rational functions of $\ell$ and a single basis function,
\begin{align}
  S_\ell
  &\equiv
  {}_3F_2\left(1,\ell,\ell;\ell+1,\ell+1;-1\right)
  =
  \frac{\ell^2}{4}
  \left[
  \psi^{(1)}\left(\frac{\ell}{2}\right)
  -\psi^{(1)}\left(\frac{\ell+1}{2}\right)
  \right].
  \label{eq:Sl_trigamma_identity}
\end{align}
Substituting this identity and simplifying the rational part gives the closed form solution
\begin{align}
  \mathcal{I}^{4,4}_\ell
  &= \frac{((\ell+4)!)^2}{1152((\ell-4)!)^2}
  \left[\psi^{(1)}\left(\frac{\ell+1}{2}\right)-\psi^{(1)}\left(\frac{\ell}{2}\right)\right]
  + R_{14}(\ell),
  \label{eq:I44_closed_form}
\end{align}
where $\psi^{(1)}$ is the trigamma function and
\begin{align}
  R_{14}(\ell)
  &= \frac{\ell^{14}}{576}+\frac{\ell^{13}}{64}-\frac{\ell^{12}}{144}-\frac{79\ell^{11}}{192}
  -\frac{193\ell^{10}}{288}+\frac{2213\ell^9}{576}+\frac{1465\ell^8}{144} \notag\\
  &\quad -\frac{8191\ell^7}{576}-\frac{3823\ell^6}{64}+\frac{2107\ell^5}{288}
  +\frac{12743\ell^4}{72}+\frac{2123\ell^3}{24} \notag\\
  &\quad -\frac{1643\ell^2}{4}-297\ell+2556.
  \label{eq:I44_polynomial}
\end{align}
The first four values in \cref{eq:I44_seed_values} are inserted separately in the implementation because the recurrence relations used for the auxiliary integrals are singular at low order.

Note that the base integral has the asymptotic expansion
\begin{gather}
  \mathcal{I}^{4,4}_\ell
  \sim \frac{40320}{(\ell+1/2)^2}
  -\frac{997920}{(\ell+1/2)^4}
  +\frac{33770520}{(\ell+1/2)^6}
  -\frac{1466324370}{(\ell+1/2)^8}
  +\cdots.
  \label{eq:I44_asymptotic}
\end{gather}
The stronger powers cancel in the full expression \cref{eq:exact_integral_final}, leaving the $\left(\ell + \frac{1}{2}\right)^{-6}$ behaviour shown in \cref{eq:Iell_asymptotic}.

\section{Derivation of the recurrence relations}\label{app:recurrences}
This appendix outlines the derivation of the recurrence relations used in \cref{sec:recursion}.  First define the unnormalised triangular integrals
\begin{gather}
  J^{m,n}_\ell
  \equiv \int_0^\infty \! \mathrm{d}q \int_0^q \! \mathrm{d}q'\,
  q^m q'^n\mathrm{e}^{-q-q'} k_\ell(q)i_\ell(q'),
  \label{eq:Jmn_appendix}
\end{gather}
so that $\mathcal{I}^{m,n}_\ell=2^{m+n+3}J^{m,n}_\ell/\pi$.  We also use
\begin{gather}
  X_\ell^{(m)} \equiv \int_0^\infty q^m \mathrm{e}^{-2q}i_\ell(q)k_\ell(q)\,\mathrm{d}q.
  \label{eq:X_appendix}
\end{gather}
The modified spherical Bessel functions obey the ladder relations
\begin{gather}
  i_{\ell+1}(z) = \left(\frac{\mathrm{d}}{\mathrm{d}z}-\frac{\ell}{z}\right)i_\ell(z)
  \quad,\quad
  k_{\ell+1}(z) = -\left(\frac{\mathrm{d}}{\mathrm{d}z}-\frac{\ell}{z}\right)k_\ell(z),
  \label{eq:bessel_ladders_appendix}
\end{gather}
and the Wronskian identity
\begin{gather}
  i_\ell'(z)k_\ell(z)-i_\ell(z)k_\ell'(z) = \frac{\pi}{2z^2},
  \label{eq:wronskian_appendix}
\end{gather}
which is the source of the inhomogeneous constants in the recurrences.

It is useful to first isolate the inner integral,
\begin{gather}
  A^{(n)}_\ell(q)\equiv
  \int_0^q \mathrm{d}q'\,q'^n\mathrm{e}^{-q'}i_\ell(q').
  \label{eq:Aell_appendix}
\end{gather}
The ladder relation for $k_{\ell+1}$ gives
\begin{align}
  J^{4,4}_{\ell+1}
  &=
  \int_0^\infty\mathrm{d}q\,q^4\mathrm{e}^{-q}
  \left[-k_\ell'(q)+\frac{\ell}{q}k_\ell(q)\right]A^{(4)}_{\ell+1}(q).
  \label{eq:J_ladder_step1}
\end{align}
Integrating the derivative term by parts gives no boundary contribution: at small $q$ the product scales as a non-negative power after the lower limit in $A^{(4)}_{\ell+1}$ is included, and at large $q$ the exponential factors dominate.  Thus
\begin{align}
  J^{4,4}_{\ell+1}
  &=
  \int_0^\infty\mathrm{d}q\,q^8\mathrm{e}^{-2q}i_{\ell+1}(q)k_\ell(q)
  +\int_0^\infty\mathrm{d}q\,(\ell+4-q)q^3\mathrm{e}^{-q}k_\ell(q)A^{(4)}_{\ell+1}(q).
  \label{eq:J_ladder_step2}
\end{align}
The first integral can be reduced to the diagonal quantities $X_\ell^{(m)}$ by using
$i_{\ell+1}=i_\ell'-\ell i_\ell/q$ and then the Wronskian:
\begin{align}
  \int_0^\infty\mathrm{d}q\,q^8\mathrm{e}^{-2q}i_{\ell+1}(q)k_\ell(q)
  &=X_\ell^{(8)}-(\ell+4)X_\ell^{(7)}+\frac{45\pi}{32}.
  \label{eq:diagonal_wronskian_appendix}
\end{align}
For the second integral, the ladder relation for $i_{\ell+1}$ gives
\begin{align}
  A^{(4)}_{\ell+1}(q)
  &=q^4\mathrm{e}^{-q}i_\ell(q)
  -\int_0^q \mathrm{d}q'\,(\ell+4-q')q'^3\mathrm{e}^{-q'}i_\ell(q').
  \label{eq:A_ladder_appendix}
\end{align}
Substituting this in \cref{eq:J_ladder_step2}, the $X_\ell^{(8)}-(\ell+4)X_\ell^{(7)}$ contribution cancels against the diagonal part of the second integral.  The remaining triangular terms are
\begin{gather}
  J^{4,4}_{\ell+1}
  =-(\ell+4)^2J^{3,3}_\ell
  +(\ell+4)\left(J^{3,4}_\ell+J^{4,3}_\ell\right)
  -J^{4,4}_\ell+\frac{45\pi}{32}.
  \label{eq:J_ladder_appendix}
\end{gather}
To close this relation one needs identities among the lower-power integrals.  These follow from the differential equation
\begin{gather}
  \left[\frac{1}{q^2}\frac{\mathrm{d}}{\mathrm{d}q}\left(q^2\frac{\mathrm{d}}{\mathrm{d}q}\right)-\frac{\ell(\ell+1)}{q^2}-1\right]y(q)=0,
  \label{eq:bessel_ode_appendix}
\end{gather}
which is satisfied by both $i_\ell$ and $k_\ell$.  Applying this operator to the $q$ or $q'$ spherical Bessel function inside $J^{m,n}_\ell$ gives the two master equations
\begin{align}
  &\left[(m-2)(m-1)-\ell(\ell+1)\right]J^{m-2,n}_\ell
  -2(m-1)J^{m-1,n}_\ell \notag\\
  &\qquad +\frac{3m+n-4}{2}X_\ell^{(m+n-1)}-2X_\ell^{(m+n)}
  +\frac{\pi(m+n-2)!}{2^{m+n+1}}=0,
  \label{eq:master1_appendix}\\
  &\left[(n-2)(n-1)-\ell(\ell+1)\right]J^{m,n-2}_\ell
  -2(n-1)J^{m,n-1}_\ell \notag\\
  &\qquad -\frac{m+3n-4}{2}X_\ell^{(m+n-1)}+2X_\ell^{(m+n)}
  +\frac{\pi(m+n-2)!}{2^{m+n+1}}=0.
  \label{eq:master2_appendix}
\end{align}
Adding \eqref{eq:master1_appendix} and \eqref{eq:master2_appendix} with interchanged indices cancels the $X$ terms.  The choices $(m,n)=(5,4)$ and $(4,5)$ give
\begin{gather}
  J^{3,4}_\ell+J^{4,3}_\ell
  =
  \frac{16}{(\ell+4)(3-\ell)}J^{4,4}_\ell
  -\frac{315\pi}{32(\ell+4)(3-\ell)},
  \label{eq:J34_J43_reduction_appendix}
\end{gather}
while $(m,n)=(5,3)$ and $(3,5)$ give
\begin{gather}
  J^{3,3}_\ell
  =
  \frac{64}{(\ell+4)^2(3-\ell)^2}J^{4,4}_\ell
  +\frac{45\pi\left(\ell^2+\ell-40\right)}{32(\ell+4)^2(3-\ell)^2}.
  \label{eq:J33_reduction_appendix}
\end{gather}
Substitution in \cref{eq:J_ladder_appendix} then leaves a single first-order recurrence,
\begin{gather}
  J^{4,4}_{\ell+1}
  =
  -\left(\frac{\ell+5}{\ell-3}\right)^2J^{4,4}_\ell
  +\frac{315\pi}{8(\ell-3)^2}.
  \label{eq:J44_recurrence_appendix}
\end{gather}
After the normalisation $\mathcal{I}^{m,n}_\ell=2^{m+n+3}J^{m,n}_\ell/\pi$, this is \cref{eq:I44_recurrence}.  Applying the same master equations to $(m,n)=(4,4)$ gives the auxiliary reduction \cref{eq:I42I24_reduction}, and \cref{eq:J33_reduction_appendix} gives \cref{eq:I33_reduction}.  The relations with $m$ or $n<3$ are evaluated directly because the closed reductions above become singular at low order.

\section{Derivation of the angular weights}\label{app:angular_weights}
This appendix derives the coefficients $W^{(i)}_{\ell k}$ defined in \cref{eq:W_definition}.  Throughout we use
\begin{align}
  u_\ell(\mu)&=(1-\mu)^2P_\ell(\mu), \\
  O^{(1)}_\ell(\mu)&=u_\ell(\mu), \\
  O^{(2)}_\ell(\mu)&=4u_\ell''(\mu), \\
  O^{(3)}_\ell(\mu)&=4(1-\mu)u_\ell''(\mu),
\end{align}
and we derive the three families in turn.

\subsection{The banded coefficients $W^{(1)}_{\ell k}$}
Since
\begin{gather}
  (1-\mu)^2=\frac{4}{3}P_0(\mu)-2P_1(\mu)+\frac{2}{3}P_2(\mu),
  \label{eq:one_minus_mu_sq_appendix}
\end{gather}
the operator $O^{(1)}_\ell$ reduces to products of $P_\ell$ with the first three Legendre polynomials.  Using the three-term recurrence\footnote{DLMF equation~14.10.3.}
\begin{gather}
  \mu P_\ell(\mu)=\frac{\ell+1}{2\ell+1}P_{\ell+1}(\mu)+\frac{\ell}{2\ell+1}P_{\ell-1}(\mu),
  \label{eq:muPell_appendix}
\end{gather}
twice gives
\begin{align}
  (1-\mu)^2P_\ell(\mu)
  &= \frac{(\ell+1)(\ell+2)}{(2\ell+1)(2\ell+3)(2\ell+5)}P_{\ell+2}(\mu)
  - \frac{2(\ell+1)}{(2\ell+1)(2\ell+3)}P_{\ell+1}(\mu) \notag\\
  &\quad + \frac{2(3\ell^2+3\ell-2)}{(2\ell-1)(2\ell+1)(2\ell+3)}P_\ell(\mu)
  - \frac{2\ell}{(2\ell-1)(2\ell+1)}P_{\ell-1}(\mu) \notag\\
  &\quad + \frac{\ell(\ell-1)}{(2\ell-3)(2\ell-1)(2\ell+1)}P_{\ell-2}(\mu).
  \label{eq:uell_banded_appendix}
\end{align}
Projecting \cref{eq:uell_banded_appendix} with \cref{eq:W_definition} immediately gives \cref{eq:W1_cases}.  In particular, $W^{(1)}_{\ell k}$ vanishes unless $k\in\{\ell-2,\ell-1,\ell,\ell+1,\ell+2\}$.

\subsection{The Legendre expansion of $u_\ell''(\mu)$}
Define
\begin{gather}
  L^{(1)}_\ell(\mu)\equiv (1-\mu)P'_\ell(\mu),
  \qquad
  L^{(2)}_\ell(\mu)\equiv (1-\mu)^2P''_\ell(\mu).
\end{gather}
Then
\begin{gather}
  u_\ell''(\mu)=2P_\ell(\mu)-4L^{(1)}_\ell(\mu)+L^{(2)}_\ell(\mu).
  \label{eq:uellpp_split_appendix}
\end{gather}
The key point is that $L^{(1)}_\ell$ and $L^{(2)}_\ell$ obey certain recurrences.  From the adjacent-order derivative relations\footnote{The first line is DLMF equation~14.10.5, while the second follows by combining DLMF equations~14.10.3 and 14.10.5 with $\ell\to\ell-1$.}
\begin{align}
  (1-\mu^2)P'_\ell(\mu) &= \ell\left[P_{\ell-1}(\mu)-\mu P_\ell(\mu)\right],
  \label{eq:adjacent_legendre_1_appendix}\\
  (1-\mu^2)P'_{\ell-1}(\mu) &= \ell\left[\mu P_{\ell-1}(\mu)-P_\ell(\mu)\right],
  \label{eq:adjacent_legendre_2_appendix}
\end{align}
we obtain
\begin{gather}
  L^{(1)}_\ell(\mu)+L^{(1)}_{\ell-1}(\mu)=\ell\left[P_{\ell-1}(\mu)-P_\ell(\mu)\right].
  \label{eq:L1_recurrence_appendix}
\end{gather}
Iterating \cref{eq:L1_recurrence_appendix} down to $L^{(1)}_0=0$ gives the finite Legendre sum
\begin{gather}
  L^{(1)}_\ell(\mu)
  =
  -\ell P_\ell(\mu)
  +\sum_{k=0}^{\ell-1}(-1)^{\ell-k-1}(2k+1)P_k(\mu).
  \label{eq:L1_closed_appendix}
\end{gather}
Differentiating \cref{eq:L1_recurrence_appendix} once more and multiplying by $(1-\mu)$ yields
\begin{gather}
  L^{(2)}_\ell(\mu)
  =
  (1-\ell)L^{(1)}_\ell(\mu)
  +\sum_{k=1}^{\ell-1}(-1)^{\ell-k-1}(2k+1)L^{(1)}_k(\mu).
  \label{eq:L2_closed_appendix}
\end{gather}
Substituting \cref{eq:L1_closed_appendix} and \cref{eq:L2_closed_appendix} into \cref{eq:uellpp_split_appendix} and collecting equal Legendre orders gives
\begin{align}
  u_\ell''(\mu)
  &= (\ell+1)(\ell+2)P_\ell(\mu) + \sum_{k=0}^{\ell-1}(-1)^{\ell-k}(2k+1)
  \left[\ell(\ell+1)-k(k+1)+2\right]P_k(\mu).
  \label{eq:uellpp_closed_appendix}
\end{align}
This explains why $W_{\ell k}^{(2)}$ and $W_{\ell k}^{(3)}$ in \cref{eq:W2_cases,eq:W3_cases} are not banded.

\subsection{The coefficients $W^{(2)}_{\ell k}$ and $W^{(3)}_{\ell k}$}
Multiplying \cref{eq:uellpp_closed_appendix} by $4$ and comparing with \cref{eq:W_definition} immediately gives \cref{eq:W2_cases},
\begin{gather}
  O^{(2)}_\ell(\mu)
  =
  4(\ell+1)(\ell+2)P_\ell(\mu)
  + 4\sum_{k=0}^{\ell-1}(-1)^{\ell-k}(2k+1)
  \left[\ell(\ell+1)-k(k+1)+2\right]P_k(\mu).
  \label{eq:O2_appendix}
\end{gather}
For $O^{(3)}_\ell$, we multiply \cref{eq:uellpp_closed_appendix} by $(1-\mu)$ and use \cref{eq:muPell_appendix} term by term.  Writing
\begin{gather}
  u_\ell''(\mu)=\sum_{k=0}^{\ell}C_{\ell k}P_k(\mu),
  \label{eq:C_definition_appendix}
\end{gather}
with
\begin{gather}
  C_{\ell\ell}=(\ell+1)(\ell+2),
  \qquad
  C_{\ell k}=(-1)^{\ell-k}(2k+1)\left[\ell(\ell+1)-k(k+1)+2\right]
  \quad (k<\ell),
  \label{eq:C_cases_appendix}
\end{gather}
the coefficient of $P_k$ in $(1-\mu)u_\ell''$ is
\begin{gather}
  C_{\ell k}
  -\frac{k}{2k-1}C_{\ell,k-1}
  -\frac{k+1}{2k+3}C_{\ell,k+1}.
  \label{eq:O3_projection_appendix}
\end{gather}
For $k<\ell-1$, substituting \cref{eq:C_cases_appendix} into \cref{eq:O3_projection_appendix} gives
\begin{align}
  &C_{\ell k}
  -\frac{k}{2k-1}C_{\ell,k-1}
  -\frac{k+1}{2k+3}C_{\ell,k+1} \notag\\
  &\qquad =
  2(-1)^{\ell-k}(2k+1)\left[\ell(\ell+1)-k(k+1)+1\right].
  \label{eq:O3_tail_appendix}
\end{align}
The highest orders must be treated separately because one or both neighbouring coefficients are absent. This gives
\begin{align}
  [P_{\ell+1}](1-\mu)u_\ell'' &= -\frac{(\ell+1)^2(\ell+2)}{2\ell+1}, \notag\\
  [P_{\ell}](1-\mu)u_\ell'' &= (\ell+1)(3\ell+2), \notag\\
  [P_{\ell-1}](1-\mu)u_\ell'' &= -\frac{17\ell^3+7\ell^2-4\ell-2}{2\ell+1},
  \label{eq:O3_top_appendix}
\end{align}
where $[P_k]$ denotes the coefficient of $P_k$.  Multiplying by the overall factor of four in $O^{(3)}_\ell$ and comparing with \cref{eq:W_definition} reproduces \cref{eq:W3_cases}.  In particular, \cref{eq:O3_tail_appendix} shows explicitly that the lower limit remains $k=0$.

\section{Implementation details and numerical approximation}\label{app:implementation_details}
The exact evaluation of \cref{eq:exact_integral_final} is straightforward:
\begin{enumerate}
  \item Generate $\mathcal{I}^{4,4}_k=r_k+s_k\pi^2$ from \eqref{eq:I44_seed_values} and \eqref{eq:I44_recurrence} up to $k=\ell+2$.
  \item Evaluate the rational weights $W^{(i)}_{\ell k}$ and the rational coefficients $a_k,b_k,c_k,d_k$.
  \item Accumulate separately the rational coefficient of $1$ and the rational coefficient of $\pi^2$ in \cref{eq:exact_integral_final}.
\end{enumerate}
To conserve memory for large $\ell$ computations, it is advantageous to roll up the recurrence inside the summation, and this is the procedure implemented in \integralnotebook.  The use of exact rational arithmetic is important, because the final double-precision value is often obtained after a large cancellation between the two exact pieces.

For the final conversion to floating point, the implementation approximates $\pi^2$ by the Chudnovsky series~\cite{Chudnovsky1989}.  With
\begin{gather}
  S_K = \sum_{j=0}^{K}\frac{(-1)^j(6j)!(545140134j+13591409)}{(3j)!(j!)^3 640320^{3j}},
  \label{eq:chudnovsky_sum}
\end{gather}
we use
\begin{gather}
  \pi^2_K = \frac{1823176476672000}{S_K^2}.
  \label{eq:pi2_chudnovsky}
\end{gather}
Terms are added until the resulting double-precision value of $I_\ell$ no longer changes.  This avoids choosing a fixed decimal precision that may be insufficient at high multipole.

We worked out the large-$\ell$ behaviour numerically, and we found that the complete kernel satisfies
\begin{gather}
  I_\ell \sim
  \frac{5529600}{(\ell+1/2)^6}
  -\frac{656640000}{(\ell+1/2)^8}
  +\frac{64159257600}{(\ell+1/2)^{10} }
  +\cdots .
  \label{eq:Iell_asymptotic}
\end{gather}
Note how the leading $\ell^{-6}$ scaling is different from the one coming from the base integral where $\mathcal{I}^{4,4}_\ell\sim \ell^{-2}$.  It emerges only after the angular weights in \cref{eq:exact_integral_final} cancel the stronger terms.

We can use the asymptotic behaviour to build a constrained rational approximation. Defining $z=(\ell+1/2)^{-2}$,
\begin{gather}
  I_\ell^{\mathrm{fit}}(z)=z^3\frac{p_0+p_1z+p_2z^2+p_3z^3+p_4z^4+p_5z^5}{1+q_1z+q_2z^2+q_3z^3+q_4z^4+q_5z^5},
  \label{eq:rational_fit}
\end{gather}
with
\begin{align}
  p_3 &= 17894987354.00044,
  &q_1 &= 265.1039577689589,
  &q_4 &= 8041726.168424849, \notag\\
  p_4 &= -9680376946.240694,
  &q_2 &= 22154.177207573077,
  &q_5 &= 21122401.83355329, \\
  p_5 &= 1623310849.867793,
  &q_3 &= 687011.2661763201, \notag
\end{align}
where the asymptotic coefficients fix
\begin{align}
  p_0 &= 5529600,
  \notag\\
  p_1 &= -656640000 + 5529600 q_1,
  \label{eq:rational_fit_constraints}\\
  p_2 &= 64159257600 - 656640000 q_1 + 5529600 q_2.
  \notag
\end{align}
The error of this approximation is below $10^{-6}$ for all $\ell \geq 0$ and goes to zero asymptotically.


\bibliographystyle{utcaps}
\newpage
\bibliography{bibliography}

\end{document}